\documentclass[%
 reprint,
 amsmath,amssymb,
 aps,nofootinbib
 ,superscriptaddress
 ]{revtex4-2}

\usepackage{hyperref}

\usepackage[utf8]{inputenc}

\usepackage{graphicx}
\usepackage{dcolumn}
\usepackage{bm}

\newcommand{\ket}[1]{\ensuremath{| {#1} \rangle }}
\newcommand{\bra}[1]{\ensuremath{\langle {#1} |}}

\renewcommand{\vec}[1]{\bm{#1}}

\newcommand{\Myquote}[1]{``#1''}

\begin{document}

\title{Teaching to extract spectral densities from lattice correlators\\ to a broad audience of learning-machines}

\author{Michele Buzzicotti}
\email[]{michele.buzzicotti@roma2.infn.it}
\affiliation{
	University and INFN of Roma Tor Vergata, Via della Ricerca Scientifica 1, I-00133, Rome, Italy
}

\author{Alessandro De Santis}
\email[]{alessandro.desantis@roma2.infn.it}
\affiliation{
	University and INFN of Roma Tor Vergata, Via della Ricerca Scientifica 1, I-00133, Rome, Italy
}

\author{Nazario Tantalo}
\email[]{nazario.tantalo@roma2.infn.it}
\affiliation{
	University and INFN of Roma Tor Vergata, Via della Ricerca Scientifica 1, I-00133, Rome, Italy
}

\date{\today}

\begin{abstract}
We present a new supervised deep-learning approach to the problem of the extraction of smeared spectral densities from Euclidean lattice correlators. A distinctive feature of our method is a model-independent training strategy that we implement by parametrizing the training sets over a functional space spanned by Chebyshev polynomials. The other distinctive feature is a reliable estimate of the systematic uncertainties that we achieve by introducing several ensembles of machines, the broad audience of the title. By training an ensemble of machines with the same number of neurons over training sets of fixed dimensions and complexity, we manage to provide a reliable estimate of the systematic errors by studying numerically the asymptotic limits of infinitely large networks and training sets. The method has been validated on a very large set of random mock data and also in the case of lattice QCD data. We extracted the strange-strange connected contribution to the smeared $R$-ratio from a lattice QCD correlator produced by the ETM Collaboration and compared the results of the new method with the ones previously obtained with the HLT method by finding a remarkably good agreement between the two totally unrelated approaches.
\end{abstract}

\nopagebreak
\maketitle

\section{Introduction}
\label{sec:intro}
The problem of the extraction of hadronic spectral densities from Euclidean correlators, computed from numerical lattice QCD simulations, has attracted a lot of attention since many years (see Refs.~\cite{Barata:1990rn,Jarrell:1996rrw,Nakahara:1999vy,Asakawa:2000tr,Aarts:2002cc,Aarts:2005hg,Laine:2008cf,Meyer:2008gt,Burnier:2013nla,Meyer:2011gj,Hansen:2017mnd,Tripolt:2018xeo,Bulava:2019kbi,Hansen:2019idp,Kades:2019wtd,Karpie:2019eiq,Kades:2019wtd,Bailas:2020qmv,Gambino:2020crt,Bruno:2020kyl,Horak:2021syv,Bulava:2021fre,Wang:2021cqw,Chen:2021jey,Chen:2021giw,Wang:2021jou,Zhou:2021bvw,Lechien:2022ieg,Boyda:2022nmh,Shi:2022yqw,Bergamaschi:2023xzx}, the works on the subject of which we are aware of, and Refs.~\cite{Rothkopf:2022fyo,Bulava:2023mjc} for recent reviews). At zero temperature, the theoretical and phenomenological importance of hadronic spectral densities, strongly emphasized in the context of lattice field theory in Refs.~\cite{Barata:1990rn,Hansen:2017mnd,Bulava:2019kbi,Hansen:2019idp,Bailas:2020qmv,Gambino:2020crt,Bruno:2020kyl}, is associated with the fact that from their knowledge it is possible to extract all the information needed to study the scattering of hadrons and, more generally, their interactions.

From the mathematical perspective, the problem of the extraction of spectral densities from lattice correlators is equivalent to that of an inverse Laplace-transform operation, to be performed numerically by starting from a discrete and finite set of noisy input data. This is a notoriously ill-posed numerical problem that, in the case of lattice field theory correlators, gets even more complicated because lattice simulations have necessarily to be performed on finite volumes where the spectral densities are badly-behaving distributions. 

\begin{figure}[t!]
\begin{center}	
\includegraphics[width=\columnwidth]{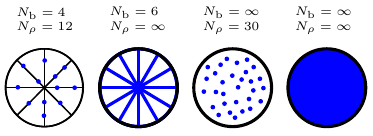}
\caption{\small 
By introducing a discrete functional-basis, with elements $B_n(E)$, that is dense in the space of square-integrable functions $f(E)$ in the interval $[E_0,\infty)$ with $E_0>0$, any such function can exactly be represented as $f(E)=\sum_{n=0}^\infty c_n B_n(E)$. With an infinite number of basis functions ($N_\mathrm{b}=\infty$) and by randomly selecting an infinite number ($N_\rho=\infty$) of coefficient vectors $\vec c=(c_0,\cdots, c_{N_\mathrm{b}})$, one can get any possible spectral density. This is the situation represented by the filled blue disk. If the number of basis functions $N_\mathrm{b}$ and the number of randomly extracted spectral densities $N_\rho$ are both finite one has a training set that is finite and that also depends on $N_\mathrm{b}$. This is the situation represented in the first disk on the left. The other two disks schematically represent the situations in which either $N_\mathrm{b}$ or $N_\rho$ is infinite.
}
\label{fig:cartoon1}
\end{center}
\end{figure}
In Ref.~\cite{Hansen:2019idp}, together with M.~Hansen and A.~Lupo, one of the authors of the present paper proposed a method to cope with the problem of the extraction of spectral densities from lattice correlators that allows to take into account the fact that distributions have to be smeared with sufficiently well-behaved test functions. Once smeared, finite volume spectral densities become numerically manageable and the problem of taking their infinite volume limit is mathematically well defined. 
The method of Ref.~\cite{Hansen:2019idp} (HLT method in short) has been further refined in Ref.~\cite{Bulava:2021fre} where it has been  validated by performing very stringent tests within the two-dimensional $O(3)$ non-linear $\sigma$-model. 

In this paper we present a new method for the extraction of smeared spectral densities from lattice correlators that is based on a supervised deep-learning approach.

The idea of using machine-learning techniques to address the problem of the extraction of spectral densities from lattice correlators is certainly not original (see e.g.\ Ref.~\cite{Karpie:2019eiq,Kades:2019wtd,Wang:2021cqw,Chen:2021jey,Chen:2021giw,Wang:2021jou,Zhou:2021bvw,Lechien:2022ieg,Boyda:2022nmh,Shi:2022yqw}). The great potential of properly-trained deep neural networks in addressing this problem is pretty evident from the previous works on the subject. These findings strongly motivated us to develop an approach that can be used to obtain trustworthy theoretical predictions. To this end we had to address the following two pivotal questions
\begin{enumerate}
\item is it possible to devise a \emph{model independent} training strategy?

\item if such a strategy is found, is it then possible to quantify reliably, together with the statistical errors, also the unavoidable \emph{systematic uncertainties}?
\end{enumerate}

The importance of the first question can hardly be underestimated. Under the working assumption, supported by the so-called universal reconstruction theorems (see Refs.~\cite{HORNIK1989359,Goodfellow-et-al-2016,cybenko1989approximation}), that a sufficiently large neural network can perform any task, limiting either the size of the network or the information to which it is exposed during the training process means, in fact, limiting its ability to solve the problem in full generality. Addressing the second question makes the difference between providing a possibly efficient but qualitative solution to the problem and providing a scientific numerical tool to be used in order to derive theoretical predictions for phenomenological analyses.

\begin{figure}[t!]
\begin{center}	
\includegraphics[width=\columnwidth]{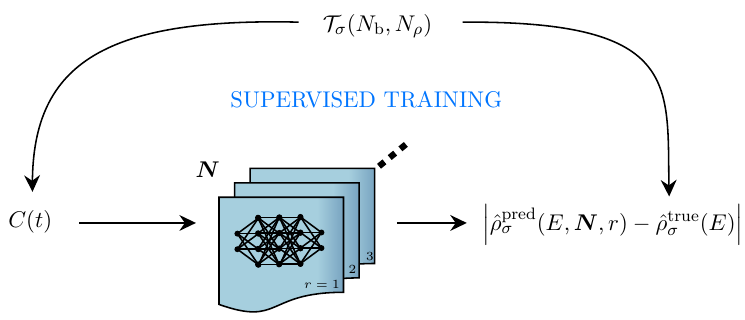}
\caption{\small  
We generate several training sets $\mathcal{T}_\sigma(N_\mathrm{b},N_\rho)$ built by considering randomly chosen spectral densities. These are obtained by choosing $N_\rho$ random coefficients vectors with $N_\mathrm{b}$ entries, see FIG.~\ref{fig:cartoon1}. For each spectral density  $\rho(E)$ we build the associated correlator $C(t)$ and smeared spectral density $\hat{\rho}_\sigma(E)$, where $\sigma$ is the smearing parameter. We then distort the correlator $C(t)$, by using the information provided by the statistical variance of the lattice correlator (the one we are going to analyse at the end of the trainings), and obtain the input-output pair $(C_\mathrm{noisy}(t),\hat{\rho}_\sigma(E))$ that we add to $\mathcal{T}_\sigma(N_\mathrm{b},N_\rho)$. We then implement different neural networks with $N_\mathrm{n}$ neurons and at fixed $\vec{N}=(N_\mathrm{n},N_\mathrm{b},N_{\rho})$ we introduce an \textit{ensemble of machines} with $N_\mathrm{r}$ replicas. Each machine $r=1,\cdots,N_\mathrm{r}$ belonging to the ensemble has the same architecture and, before the training, differs from the other replicas for the initialization parameters. All the replicas are then trained over $\mathcal{T}_\sigma(N_\mathrm{b},N_\rho)$ and, at the end of the training process, each replica will give a different answer depending upon $\vec{N}$.
}
\label{fig:cartoon2}
\end{center}
\end{figure}
\begin{figure}[t!]
\begin{center}	
\includegraphics[width=\columnwidth]{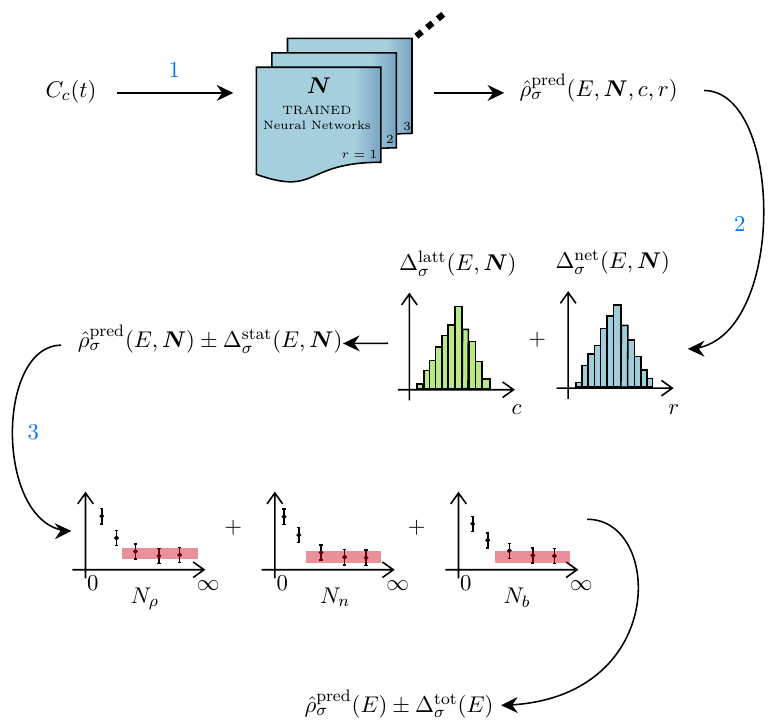}
\caption{\small
Flowchart illustrating the three-step procedure that, after the training, we use to extract the final result. Here $C(t)$ represents the input lattice correlator that, coming from a Monte Carlo simulation, is affected by statistical noise. We call $C_c(t)$ the $c$-th bootstrap sample (or jackknife bin) of the lattice correlator with $c=1,\cdots, N_\mathrm{c}$. In the first step, $C_c(t)$ is fed to all the trained neural networks belonging to the ensemble at fixed $\vec{N}$ and the corresponding  answers $\hat{\rho}_\sigma^\mathrm{pred}(E,\vec{N},c,r)$ are collected. In the second step, by combining in quadrature the widths of the distributions of the answers as a function of the index $c$ ($\Delta_\sigma^\mathrm{latt}(E,\vec{N})$) and of the index $r$ ($\Delta_\sigma^\mathrm{net}(E,\vec{N})$) we estimate the error $\Delta_\sigma^\mathrm{stat}(E,\vec{N})$) at fixed $\vec N$.  At the end of this step we are left with a collection of results $\hat{\rho}_\sigma^\mathrm{pred}(E,\vec{N})\pm \Delta_\sigma^\mathrm{stat}(E,\vec{N})$. In the third and last step,  the limits $\vec N\mapsto \infty$ are studied numerically and an unbiased estimate of $\hat{\rho}_\sigma^\mathrm{pred}(E)$ and of its error $\Delta_\sigma^\mathrm{tot}(E)$, with the latter also taking into account the unavoidable systematics associated with these limits, is finally obtained.
}
\label{fig:cartoon3}
\end{center}
\end{figure}
In order to address these two questions, the method that we propose in this paper has been built on two pillars
\begin{enumerate}
\item the introduction of a \emph{functional-basis} to parametrize the correlators and the smeared spectral densities of the training sets in a model independent way;

\item the introduction of the \emph{ensemble of machines}, the broad audience mentioned in the title, to estimate the systematic errors\footnote{The idea of using ensembles of machines has already been explored within the machine-learning literature in other contexts, see e.g.\ Ref.~\cite{pmlr-v108-pearce20a} where the so-called ``ensembling'' technique to estimate the errors is discussed from a Bayesian perspective.}.
\end{enumerate}
A bird-eye view of the proposed strategy is given in FIG.~\ref{fig:cartoon1}, FIG.~\ref{fig:cartoon2} and in FIG.~\ref{fig:cartoon3}. The method is validated by using both mock and true lattice QCD data. In the case of mock data the exact result is known and the validation test is quite stringent. In the case of true lattice QCD data the results obtained with the new method are validated by comparison with the results obtained with the HLT method.

The plan of the paper is as follows.  In Section~\ref{sec:problem} we introduce and discuss the main aspects of the problem. In Section~\ref{sec:setup} 
we illustrate the numerical setup used to obtain the results presented in the following sections.
In Section~\ref{sec:training_strategy} we describe the construction of the training sets and the proposed model-independent training strategy. In Section~\ref{sec:quote_results} we illustrate the procedure that we use to extract predictions from our ensembles of trained machines and present the results of a large number of validation tests performed with random mock data. Further validation tests, performed by using mock data generated by starting from physically inspired models, are presented in Section~\ref{sec:mock_data}. In Section~\ref{sec:lattice_data} we present our results in the case of lattice QCD data and the comparison with the HLT method. We draw our conclusions in Section~\ref{sec:conclusions} and discuss some technical details in the two appendixes.

\section{\label{sec:problem}Theoretical Framework}
The problem that we want to solve is the extraction of the infinite-volume spectral density $\rho(E)$ from the Euclidean correlators
\begin{flalign}
&
C_{LT}(t)= 
\int_{E_0}^\infty\, \mathrm{d} E\, b_T(t,E)\, \rho_{LT}(E)\;,
\label{eq:Cstarting}
\end{flalign}
computed numerically in lattice simulations. These are performed by introducing a finite spatial volume $L^3$, a finite Euclidean temporal direction $T$ and by discretizing space-time,
\begin{flalign}
x=a(\tau,\vec n)\;,
\quad
0\le \tau< N_T=\frac{T}{a}\;,
\quad
0\le n_i< N_L=\frac{L}{a}\;,
\label{eq:lattice}
\end{flalign}
where $a$ is the so-called lattice spacing and $(\tau,\vec n)$ are the integer space-time coordinates in units of $a$. In order to obtain the infinite-volume spectral density one has to perform different lattice simulations, by progressively increasing $L$ and/or $T$, and then study numerically the $L\mapsto \infty$ and $T\mapsto \infty$ limits. In the $T\mapsto \infty$ limit the basis functions $b_T(t,E)$ become decaying exponentials and the correlator is given by 
\begin{flalign}
&
C_{L}(t)=
\lim_{T\mapsto\infty} C_{TL}(t)=
\int_{E_0}^\infty\, \mathrm{d}E\, e^{-tE}\, \rho_{L}(E)\;,
\label{eq:CfiniteL}
\end{flalign}
where $\rho_{L}(E)=0$ for $E<E_0$ and the problem is that of performing a numerical inverse Laplace-transform operation, by starting from a discrete and finite set of noisy input data. This is a classical numerical problem, arising in many research fields, and has been thoroughly studied (see e.g.\ Refs.~\cite{books/lib/RasmussenW06,murphy2013machine} for textbooks discussing the subject from a machine-learning perspective). The problem, as we are now going to discuss, is particularly challenging from the numerical point of view and becomes even more challenging and delicate in our Quantum Field Theory (QFT) context because, according to Wightman's axioms, QFT spectral densities live in the space of tempered distributions and, therefore, cannot in general be considered smooth and well behaved functions.

Before discussing though the aspects of the problem that are peculiar to our QFT context, it is instructive to first review the general aspects of the numerical inverse Laplace-transform problem that, in fact,  is ill-posed in the sense of Hadamard. To this end, we start by considering the correlator in the infinite $L$ and $T$ limits, 
\begin{flalign}
&
C(a\tau)=
\lim_{L,T\mapsto\infty} C_{TL}(a\tau)=
\int_{E_0}^\infty\, \mathrm{d}E\, e^{-\tau aE}\, \rho(E)\;,
\label{eq:Cinfinite}
\end{flalign}
and we assume that our knowledge of $C(t)$ is limited to the discrete and finite set of times $t=a\tau$. Moreover we assume that the input data are affected by numerical and/or statistical errors that we call $\Delta(a\tau)$. 

The main point to be noticed is that, in general, the spectral density $\rho(E)$ contains an infinite amount of information that cannot be extracted from the limited and noisy information contained into the input data $C(a\tau)$. As a consequence, in any numerical approach to the extraction of $\rho(E)$ a discretization of Eq.~(\ref{eq:Cinfinite}) has to be introduced. Once a strategy to discretize the problem has been implemented, the resulting spectral density has then to be interpreted as a ``filtered'' or (as is more natural to call it in our context) \emph{smeared} version of the exact spectral density,
\begin{flalign}\label{eq:Krho}
\hat \rho(E)=\int_{E_0}^\infty\, \mathrm{d}\omega\, K(E,\omega)\, \rho(\omega)\;,
\end{flalign}
where $K(E,\omega)$ is the so-called smearing kernel, explicitly or implicitly introduced within the chosen numerical approach. 

There are two main strategies (and many variants of them) to discretize Eq.~(\ref{eq:Cinfinite}). The one that we will adopt in this paper has been introduced and pioneered by Backus and Gilbert~\cite{Backus} and is based on the introduction of a smearing kernel in the first place. We will call this the ``smearing'' discretization approach. The other approach, that is more frequently used in the literature and that we will therefore call the ``standard'' one, is built on the assumption that spectral densities are smooth and well-behaved functions. Before discussing the smearing approach we briefly review the standard one, by putting the emphasis on the fact that also in this case a smearing kernel is implicitly introduced.

In the standard discretization approach the infinite-volume correlator is approximated as a Riemann sum,
\begin{flalign}
&
\hat C(a\tau)=
\sigma \sum_{m=0}^{N_{E}-1}\, e^{-\tau a E_m}\, \hat \rho(E_m)\;,
\quad 
E_m=E_0+m\sigma\;,
\label{eq:CRiemann}
\end{flalign}
under the assumption that the infinite-volume spectral density is sufficiently regular to have 
\begin{flalign}
\left\vert C(a\tau) -\hat C(a\tau)\right\vert\ll \Delta(a\tau)
\end{flalign} 
for $N_E$ sufficiently large and $\sigma$ sufficiently small. By introducing the ``veilbein matrix''
\begin{flalign} 
\hat{\mathcal{E}}_{\tau m} \equiv e^{-\tau a E_m}\;,
\end{flalign} 
and the associated ``metric matrix'' in energy space,
\begin{flalign} 
&
\hat{G}_{n m} \equiv \left[\hat{\mathcal{E}}^T \hat{\mathcal{E}}\right]_{nm} 
\nonumber \\
\nonumber \\
&= \sum_{\tau=1}^{N_T-1}e^{-\tau a (E_n+E_m)} = \frac{e^{-a (E_n+E_m)}-e^{-T (E_n+E_m)}}{1-e^{-a (E_n+E_m)}}\;,
\end{flalign} 
Eq.~(\ref{eq:CRiemann}) is then solved,
\begin{flalign}
&
\hat \rho(E_n)
=
\sum_{\tau=1}^{N_T-1} g_\tau(E_n)\, \hat C(a\tau)\;,
\nonumber \\
\nonumber \\
&
g_\tau(E_n)= \frac{1}{\sigma}\sum_{m=0}^{N_{E}-1}\, \hat{G}^{-1}_{n m} \hat{\mathcal{E}}_{\tau m} \;.
\label{eq:poorman}
\end{flalign}
By using the previous expressions we can now explain why the problem is particularly challenging and in which sense it is numerically ill-posed. 

On the numerical side, the metric matrix $\hat G$ is very badly conditioned in the limit of large $N_E$ and small $\sigma$. Consequently, the coefficients $g_\tau(E_n)$ become huge in magnitude and oscillating in sign in this limit and even a tiny distortion of the input data gets enormously amplified,
\begin{flalign}
&
\sum_{\tau=1}^{N_T-1} g_\tau(E_n)\, \Delta(a\tau)\quad 
\stackrel{\sigma\mapsto 0}{\longmapsto}
\quad \infty\;.
\end{flalign}
In this sense the numerical solution becomes unstable and the problem ill-posed. 

\begin{figure}
\includegraphics[width=\columnwidth]{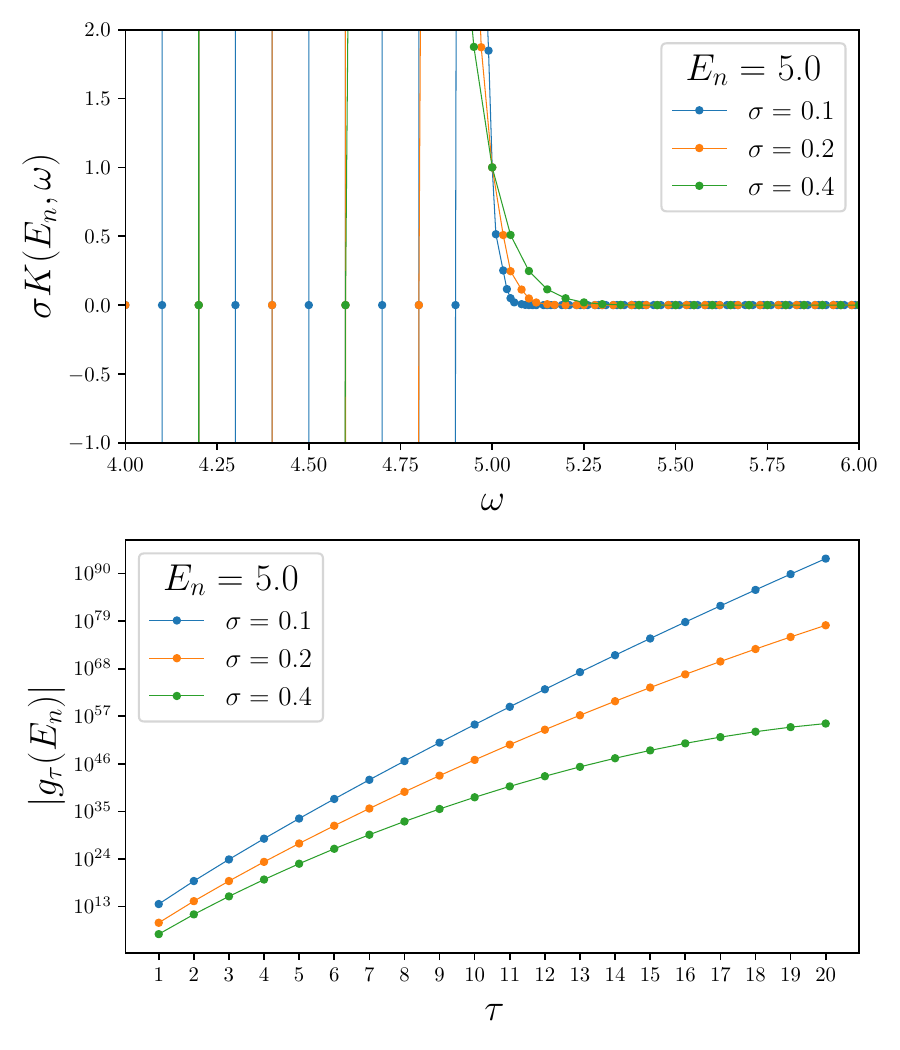}
\caption{\small \emph{Top panel}: Smearing kernel of Eq.~\ref{eq:poorman2} at $E_n=5$ for three values of $\sigma$. The reconstruction is performed by setting $E_0=1$ and $E_\mathrm{max}=10$ as UV cutoff in Eq.~\ref{eq:Krho} and by working in lattice units with  $a=1$. The kernel is smooth for $\omega \ge E_n$ while it presents huge oscillations for $\omega<E_n$. In the case $\sigma=0.1$ these oscillations range from $-10^{126}$ to $+10^{121}$.
	\emph{Bottom panel}: Corresponding coefficients $g_\tau(E_n)$ (see Eq.~\ref{eq:poorman}) in absolute value for the first 20 discrete times.
	Extended-precision arithmetic is mandatory to invert the matrix $\hat{G}_{nm}$. This test, in which $\det(\hat{G})=\mathcal{O}\big(10^{-15700}\big)$ for $\sigma=0.1$, has been performed by using 600-digits arithmetic
}
\label{fig:Kfilter}
\end{figure}

Another important observation concerning Eqs.~(\ref{eq:poorman}), usually left implicit, concerns the interpretation of $\hat \rho(E_n)$ as a smeared spectral density. By introducing the smearing kernel
\begin{flalign}
&
K(E_n,\omega)
=
\sum_{\tau=1}^{N_T-1} g_\tau(E_n)\, e^{-a\tau\omega}\;,
\label{eq:poorman2}
\end{flalign}
and by noticing that, as a matter of fact, the spectral density has to be obtained by using the correlator $C(a\tau)$ (and not its approximation $\hat C(a\tau)$, to be considered just as a theoretical device introduced in order to formalize the problem), we have
\begin{flalign}
&
\hat \rho(E_n)
=
\int_{E_0}^\infty\, \mathrm{d}\omega\, K(E_n,\omega)\, \rho(\omega)\;.
\label{eq:poormansmeared}
\end{flalign}
Consistency would require that $K(E_n,\omega)=\delta(E_n-\omega)$ but this cannot happen at finite $T$ and/or $N_E$. In fact $K(E_n,\omega)$ can be considered a numerical approximation of $\delta(E_n-\omega)$ that, as can easily be understood by noticing that
\begin{flalign}
&
K(E_n,E_m)
=
\frac{\delta_{nm}}{\sigma}\;, \label{eq:poorman3}
\end{flalign}
has an intrinsic energy-resolution proportional to the discretization interval $\sigma$ of the Riemann's sum. A numerical study of $K(E_n,\omega)$ at fixed $E_n$ reveals that for $\omega\ge E_n$ the kernel behaves smoothly while it oscillates wildly for $\omega< E_n$ and small values of $\sigma$. A numerical example is provided in FIG.~\ref{fig:Kfilter}.

Once the fact that a smearing operation is unavoidable in any numerical approach to the inverse Laplace-transform problem has been recognized, the smearing discretization approach that we are now going to discuss appears more natural. Indeed, the starting point of the smearing approach is precisely the introduction of a smearing kernel and this allows to cope with the problem also in the case, relevant for our QFT applications, in which spectral densities are distributions. 

By reversing the logic of the standard approach, that led us to Eq.~(\ref{eq:poorman2}), in the smearing approach the problem is discretized by representing the possible smearing kernels as functionals of the coefficients $g_\tau$ according to
\begin{flalign}
&
K(\vec g,\omega)
=
\sum_{\tau=1}^{N_T-1} g_\tau\, b_T(a\tau,\omega)\;.
\label{eq:BGdisctretization1}
\end{flalign}
Notice that we are now considering the correlator $C_{LT}(a\tau)$ at finite $T$ and $L$ and this allows to analyse the results of a single lattice simulation. The main observation is that in the $T\mapsto \infty$ limit any target kernel $K_\mathrm{target}(E,\omega)$ such that
\begin{flalign}
\left\|K_\mathrm{target}(E) \right\|^2= \int_{E_0}^\infty\, \mathrm{d}\omega\, \left\vert K_\mathrm{target}(E,\omega) \right\vert^2 < \infty,
\end{flalign}
can exactly be represented as a linear combination of the $b_\infty(a\tau,\omega)=\exp(-a\omega \tau)$ basis functions (that are indeed dense in the functional-space $L_2[E_0,\infty]$ of square-integrable functions). With a finite number of lattice times, the smearing kernel is defined as the best possible approximation of $K_\mathrm{target}(E,\omega)$, i.e.\ the coefficients $g_\tau(E)$ are determined by the condition
\begin{flalign}
\left.\frac{\partial \left\|K(\vec g) -K_\mathrm{target}(E) \right\|^2}{\partial g_\tau}\right\vert_{g_\tau=g_\tau(E)}=0\;.
\label{eq:BGcoefficients}
\end{flalign}
Once the coefficients are given, the smeared spectral density is readily computed by relying on the linearity of the problem,
\begin{flalign}
&
\hat \rho_{LT}(E)
=
\sum_{\tau=1}^{N_T-1} g_\tau(E)\, C_{LT}(a\tau)
\nonumber \\
\nonumber \\
&=
\int_{E_0}^\infty\, \mathrm{d}\omega \, K(E,\omega)\, \rho_{LT}(\omega)\;,
\label{eq:BGsolution}
\end{flalign}
where now
\begin{flalign}
&
K(E,\omega)\equiv K(\vec g(E),\omega)\;.
\end{flalign}
In the smearing approach one has
\begin{flalign}
&
\lim_{T\mapsto \infty}\hat \rho_{LT}(E)
=
\int_{E_0}^\infty\, \mathrm{d} \omega\, K_\mathrm{target}(E,\omega)\, \rho_{L}(E)\;.
\label{eq:BGsolution2}
\end{flalign}
Moreover, provided that $\rho(E)$ exists, it can be obtained by choosing as the target kernel a smooth approximation to $\delta(E-\omega)$ depending upon a resolution parameter $\sigma$, e.g. 
\begin{flalign}\label{eq:gaussian}
&
K_\mathrm{target}(E,\omega) = \frac{1}{\sqrt{2\pi} \sigma} e^{-\frac{(E-\omega)^2}{2\sigma^2}}\;,
\nonumber \\
\nonumber \\
&
\lim_{\sigma\mapsto 0}K_\mathrm{target}(E,\omega) = \delta(E-\omega)\;,
\end{flalign}
and by considering the limits
\begin{flalign}
&
\lim_{\sigma\mapsto 0}\lim_{L,T\mapsto \infty} \hat \rho_{LT}(E)
=
\rho(E)\;,
\label{eq:BGsolution3}
\end{flalign}
in the specified order.
When $\sigma$ is very small, the coefficients $g_\tau(E)$ determined by using Eq.~(\ref{eq:BGcoefficients}) become gigantic in magnitude and oscillating in sign, as in the case of the coefficients obtained by using Eq~(\ref{eq:poorman}). In fact the numerical problem is ill-posed independently of the approach used to discretize it.

We now come to the aspects of the problem that are peculiar to our QFT context. Hadronic spectral densities 
\begin{flalign}
&
\rho(p_1,\cdots p_{n-1})
\nonumber \\
\nonumber \\
&
=\bra{0}\hat O_1 (2\pi)^3 \delta^{4}(\hat P-p_1)\hat O_2\cdots (2\pi)^3 \delta^{4}(\hat P-p_{n-1}) \hat O_n\ket{0}\;,
\end{flalign}
are the Fourier transforms of Wightman's functions in Minkowski space, 
\begin{flalign}
&
W(x_1,\cdots x_{n-1})=\bra{0}\hat O_1 e^{-i\hat P\cdot x_1} \hat O_2\cdots e^{-i\hat P\cdot x_{n-1}} \hat O_n\ket{0}\;,
\end{flalign}
where $\hat P=(\hat H, \vec{\hat P})$ is the QCD four-momentum operator and the $\hat O_i$'s are generic hadronic operators. According to Wightman's axioms, vacuum expectation values of field operators are tempered distributions. This implies that also their Fourier transforms, the spectral densities, are tempered distributions in energy-momentum space. 
It is therefore impossible, in general, to compute numerically a spectral density. On the other hand, it is always possible to compute \emph{smeared} spectral densities,
\begin{flalign}
&
\rho[K_1,\cdots K_{n-1}]=\int \left\{\prod_i \mathrm{d}^4p_i\, K_i(p_i)\right\}\, \rho(p_1,\cdots p_{n-1})\;,
\end{flalign}
where the kernels $K_i(p)$ are Schwartz functions. 

In this paper, to simplify the discussion and the notation, we focus on the dependence upon a single space-time variable,
\begin{flalign}
&
W(x)=\bra{0}\hat O_F\, e^{-i\hat P\cdot x} \hat O_I\ket{0}\;,
\end{flalign}
where the operators $\hat O_I$ and $\hat O_F$ might also depend upon other coordinates. The spectral density associated with $W(x)$ is given by
\begin{flalign}
&
\rho(E)=\frac{1}{2\pi}\int \mathrm{d}^4 x\, e^{ip\cdot x}\, W(x)
\nonumber \\
\nonumber \\
&
=\bra{0}\hat O_F\, \delta\left(\hat H -E \right) (2\pi)^3\delta^{3}\left(\vec{\hat P} -\vec p \right)\hat O_I\ket{0}\;,
\label{eq:defrho}
\end{flalign}
and, to further simplify the notation, we don't show explicitly the dependence of $\rho(E)$ w.r.t.\ the fixed spatial momentum $\vec p$. The very same spectral density appears in the two-point Euclidean correlator
\begin{flalign}
&
C(t)=\int \mathrm{d}^3x\, e^{-i\vec p\cdot \vec x }\bra{0}\hat O_F\, e^{-t\hat H +i\vec{\hat P}\cdot \vec x} \hat O_I\ket{0}
\nonumber \\
\nonumber \\
&=
\int_{E_0}^\infty\, \mathrm{d}E\, e^{-tE}\, \rho(E)
\end{flalign}
for positive Euclidean time $t$. In the last line of the previous equation we have used the fact that, because of the presence of the energy Dirac-delta function appearing in Eq.~(\ref{eq:defrho}), the spectral density vanishes for $E<E_0$ where $E_0$ is the smallest energy that a state propagating between the two hadronic operators $\hat O_I$  and $\hat O_F$ can have. 

On a finite volume the spectrum of the Hamiltonian is discrete and, consequently, the finite-volume spectral density $\rho_L(E)$ becomes a particularly wild distribution, 
\begin{flalign}\label{eq:wilddistrib}
\rho_L(E) = \sum_{n} w_n(L)\, \delta(E_n(L)-E)\;,
\end{flalign}
the sum of isolated Dirac-delta peaks in correspondence of the eigenvalues $E_n(L)$ of the finite volume Hamiltonian. By using the previous expression one has that
\begin{flalign}
&
C_L(t)= \sum_{n} w_n(L)\, e^{-E_n(L) t}\;,
\end{flalign}
and this explains why we have discussed the standard approach to the discretization of the inverse Laplace-transform problem by first taking the infinite-volume limit. Indeed if the Riemann's discretization of Eq.~(\ref{eq:CRiemann}) is applied to $C_L(a\tau)$ and all the $E_m$'s are different from all the $E_n(L)$'s one gets $\hat C_L(a\tau)=0$!

On the one hand, also in infinite volume, spectral densities have to be handled with the care required to cope with tempered distributions and this \emph{excludes} the option of using the standard approach to discretize the problem in the first place. On the other hand, in some specific cases it might be conceivable to assume that the infinite volume spectral density is sufficiently smooth to attempt using the standard discretization approach. This requires that the infinite-volume limit of the correlator has been numerically taken and that the associated systematic uncertainty has been properly quantified (a non-trivial numerical task at small Euclidean times where the errors on the lattice correlators are tiny).

The smearing discretization approach can be used either in finite- and infinite-volume and the associated systematic errors can reliably be quantified by studying numerically the limits of Eq.~(\ref{eq:BGsolution3}). For this reason, as in the case of the HLT method of Ref.~\cite{Hansen:2019idp}, the target of the machine-learning method that we are now going to discuss in details is the extraction of \emph{smeared} spectral densities.

\section{\label{sec:setup}Numerical Setup}

In order to implement the strategy sketched in FIG.~\ref{fig:cartoon1}, FIG.~\ref{fig:cartoon2}, FIG.~\ref{fig:cartoon3} and discussed in details in the following sections, the numerical setup needs to be specified. In this section we describe the data layout that we used to represent the input correlators and the output smeared spectral densities and then provide the details of the architectures of the neural networks that we used in our study.

\subsection{\label{sec:inoutformat}
Data layout}

In this work we have considered both mock and real lattice QCD data and have chosen the algorithmic parameters in order to be able to extract the hadronic spectral density from the lattice QCD correlator described in Section~\ref{sec:lattice_data}. Since in the case of the lattice QCD correlator the basis function $b_T(t,E)$ (see Eq.~(\ref{eq:Cstarting})) is given by
\begin{flalign}
b_T(t,\omega)=\frac{\omega^2}{12\pi^2}
\left[ e^{-t\omega}+e^{-(T-t)\omega}\right]\;,
\end{flalign}
with $T=64 a$, we considered the same setup also in the case of mock data. 
These are built by starting from a model unsmeared spectral density $\rho(E)$ and by computing the associated correlator according to
\begin{flalign}
C(t)=\int_{E_0}^\infty \mathrm{d}\omega 
\frac{\omega^2}{12\pi^2}
\left[ e^{-t\omega}+e^{-(T-t)\omega}\right] \rho(\omega)\;,
\label{eq:inputC}
\end{flalign}
and the associated smeared spectral density according to
\begin{flalign}
	\hat{\rho}_\sigma(E) = \int_{E_0}^{\infty} \mathrm{d} \omega \, K_\sigma(E,\omega) \rho(\omega)\;.
\label{eq:outputRho}	
\end{flalign}
Since in the case of the lattice QCD spectral density, in order to compare the results obtained with the proposed new method with the ones previously obtained in Ref.~\cite{ExtendedTwistedMassCollaborationETMC:2022sta}, we considered a Gaussian smearing kernel, also in the case of mock data we made the choice
\begin{flalign}
	K_\sigma(E,\omega) = \frac{1}{\sqrt{2\pi} \sigma} e^{-\frac{(E-\omega)^2}{2\sigma^2}}\;.
\end{flalign}
We stress that there is no algorithmic reason behind the choice of the Gaussian as the smearing kernel and that any other kernel can easily be implemented within the proposed strategy.

Among the many computational paradigms available within the machine-learning framework, we opted for the most direct one and represented both the correlator and the smeared spectral density as finite dimensional vectors, that we used respectively as input and output of our neural networks. More precisely, the dimension of the input correlator vector has been fixed to $N_T=64$, coinciding with the available number of Euclidean lattice times in the case of the lattice QCD correlator. The inputs of the neural networks are thus the 64-components vectors  $\vec C=\{C(a),C(2a),\cdots,C(64a)\}$. The output vectors are instead given by $\vec{\hat{\rho}_\sigma}=\{\hat{\rho}_\sigma(E_\mathrm{min}),\cdots,\hat{\rho}_\sigma(E_\mathrm{max})\}$. As in Ref.~\cite{ExtendedTwistedMassCollaborationETMC:2022sta}, we have chosen to measure energies in units of the muon mass, $m_\mu=0.10566$~GeV, and set $E_\mathrm{min}=m_\mu$ and $E_\mathrm{max}=24m_\mu$. The interval $[E_\mathrm{min},E_\mathrm{max}]$ has been discretized in steps of size $m_\mu/2$. With these choices our output smeared spectral densities are the vectors $\vec{\hat{\rho}_\sigma}$ with $N_E=47$ elements corresponding to energies ranging from about $100$~MeV to $2.5$~GeV.

The noise-to-signal ratio in (generic) lattice QCD correlators increases exponentially at large Euclidean times. For this reason it might be numerically convenient to choose $N_T<T/a$ and discard the correlator at large times where the noise-to-signal ratio is bigger than one. According to our experience with the HLT method, that inherits the original Backus-Gilbert regularization mechanism, it is numerically inconvenient to discard part of the information available on the correlator provided that the information contained in the noise is used to conveniently regularize the numerical problem. Also in this new method, as we are going to explain in subsection~\ref{sec:noise}, we use the information contained in the noise of the lattice correlator during the training process and this, as shown in Appendix~\ref{sec:input_time_slices}, allows us to conveniently use all the available information on the lattice correlator, i.e.\ to set $N_T=T/a$, in order to extract the smeared spectral density. 

We treat $\sigma$, the width of the smearing Gaussian, as a fixed parameter by including in the corresponding training sets only spectral functions that are smeared with the chosen value of $\sigma$. This is a rather demanding numerical strategy because in order to change $\sigma$ the neural networks have to be trained anew, by replacing the smeared spectral densities in the training sets with those corresponding to the new value of $\sigma$.  
Architectures that give the possibility to take into account a variable input parameter, and a corresponding variable output at fixed input vector, have been extensively studied in the machine learning literature and we leave a numerical investigation of this option to future work on the subject. In this work we considered two different values, $\sigma=0.44$~GeV and $\sigma=0.63$~GeV, that correspond respectively to the smallest and largest values used in Ref.~\cite{ExtendedTwistedMassCollaborationETMC:2022sta}.

\subsection{Architectures}
By reading the discussion on the data layout presented in the previous subsection from the machine-learning point of view, we are in fact implementing neural networks to solve a $\mathbb{R}^{N_T}\mapsto \mathbb{R}^{N_E}$ regression problem with $N_T=64$ and $N_E=47$ which, from now on, fix the dimension of the input and output layers of the neural networks. 

There are no general rules to prefer a given network architecture among the different possibilities that have been considered within the machine-learning literature and it is common practice to make the choice by taking into account the details of the problem at hand. For our analysis, after having performed a comparative study at (almost) fixed number of parameters of the so-called Multilayer perceptron and  convolutional neural networks, we used feed-forward convolutional neural networks based on the LeNet architecture introduced in  Ref.~\cite{lecun1998gradient}. 

We studied in details the dependence of the output of the neural networks upon their size $N_\mathrm{n}$ and, to this end, we implemented three architectures that we called \textit{arcS}, \textit{arcM}  and \textit{arcL}. These architectures, described in full details in TABLEs~\ref{tab:arcS}, \ref{tab:arcM} and \ref{tab:arcL}, differ only for the number of maps in the convolutional layers. The number of maps are chosen so that the number of parameters of \textit{arcS:arcM:arcL} are approximately in the proportion $1:2:3$. For the implementation and the training we employed both Keras~\cite{chollet2015keras} and TensorFlow~\cite{tensorflow2015-whitepaper}.
\begin{table}[t!]
	\begin{ruledtabular}
		\begin{tabular}{lccccl}
			Type & Maps & Size & Kernel size & Stride & Activation  \\  \hline
			Input       &      & 64     &    &    &         	 \\
			Conv1D      & 2   & 32x2    & 3  & 2  & LeakyReLu    \\
			Conv1D      & 4   & 16x4    & 3  & 2  & LeakyReLu    \\
			Conv1D      & 8   & 8x8     & 3  & 2  & LeakyReLu    \\
			Flatten     &      & 384    &    &    &              \\
			Fully conn. &      & 256    &    &    & LeakyReLu    \\
			Fully conn. &      & 256    &    &    & LeakyReLu    \\
			Output      &      & 47     &    &    &             \\     \hline
			Parameters  &  94651    &		&    &	  &
		\end{tabular}
	\end{ruledtabular}
	\caption{\label{tab:arcS} \textit{arcS}: the smallest neural network architecture used in this work. The architecture is of the type feed-forward and the structure can be read from top to bottom of the table. It consist of three 1D convolutional layers with an increasing number of maps followed by two fully connected layers. The two blocks are intermediated by one flatten layer. The column denoted by \Myquote{Size} reports the shape of the signal produced by the corresponding layer. The stride of the filters is set to 2 in such a way that the dimension of the signal is halved at each 1D convolutional layer thus favouring the neural network to learn a more abstract, and possibly more effective, representation of the input data. 	As activation functions we use the LeakyReLu with negative slope coefficient  set to $-0.2$.
	The neurons with activation functions are also provided with biases.	The output is devoid of  activation function in order not to limit the output range. The bottom line reports the total number of trainable parameters. 
	}
\end{table}
\begin{table}[t!]
	\begin{ruledtabular}
		\begin{tabular}{lccccl}
			Type & Maps & Size & Kernel size & Stride & Activation  \\  \hline
			Input       &      & 64     &    &    &         	 \\
			Conv1D      & 12   & 32x12  & 3  & 2  & LeakyReLu    \\
			Conv1D      & 24   & 16x24  & 3  & 2  & LeakyReLu    \\
			Conv1D      & 48   & 8x48  & 3  & 2  & LeakyReLu    \\
			Flatten     &      & 384   &    &    &              \\
			Fully conn. &      & 256    &    &    & LeakyReLu    \\
			Fully conn. &      & 256    &    &    & LeakyReLu    \\
			Output      &      & 47     &    &    &             \\ \hline
			Parameters  &  180871    &		&    &	  &
		\end{tabular}
	\end{ruledtabular}
	\caption{\label{tab:arcM} \textit{arcM}: the medium-size architecture used in this work. See TABLE~\ref{tab:arcS} for the description.
	}
\end{table}
\begin{table}[t!]
	\begin{ruledtabular}
		\begin{tabular}{lccccl}
			Type & Maps & Size & Kernel size & Stride & Activation  \\  \hline
			Input       &      & 64     &    &    &         	 \\
			Conv1D      & 32   & 32x32  & 3  & 2  & LeakyReLu    \\
			Conv1D      & 64   & 16x64  & 3  & 2  & LeakyReLu    \\
			Conv1D      & 128  & 8x128  & 3  & 2  & LeakyReLu    \\
			Flatten     &      & 1024   &    &    &              \\
			Fully conn. &      & 256    &    &    & LeakyReLu    \\
			Fully conn. &      & 256    &    &    & LeakyReLu    \\
			Output      &      & 47     &    &    &             \\ \hline
			Parameters  &  371311    &		&    &	  &
		\end{tabular}
	\end{ruledtabular}
	\caption{\label{tab:arcL} \textit{arcL}: the largest architecture used in this work. See TABLE~\ref{tab:arcS} for the description.
	}
\end{table}
%

\section{\label{sec:training_strategy}
Model Independent Training}

In the supervised deep-learning framework a neural network is trained over a training set which is representative of the problem that has to be solved. In our case the inputs to each neural network are the correlators $\vec{C}$ and the target outputs are the associated smeared spectral densities $\vec{\hat{\rho}_\sigma}$. As discussed in the Introduction, our main goal is to devise a model-independent training strategy. To this end, the challenge is that of building a training set which contains enough variability so that, once trained, the network is able to provide the correct answer, within the quoted  errors, for any possible input correlator. 

As a matter of fact, the situation in which the neural network can exactly reconstruct any possible function is merely ideal. That would be possible only in absence of errors on the input data and with a neural network with an infinite number of neurons, trained on an infinitely large and complex training set. This is obviously impossible and in fact our goal is the realistic task of getting an output for the smeared spectral density as close as possible to the exact one by trading the unavoidable limited abilities of the neural network with a reliable estimate of the systematic error. In order to face this challenge we used the algorithmic strategy described in FIG.~\ref{fig:cartoon1}, FIG.~\ref{fig:cartoon2} and in FIG.~\ref{fig:cartoon3}. In our strategy, 
\begin{itemize}

\item the fact that the network cannot be infinitely large is parametrized by the fact that $N_\mathrm{n}$ (the number of neurons) is finite;   

\item the fact that during the training a network cannot be exposed to any possible spectral density is parametrized by the fact that $N_\mathrm{b}$ (the number of basis functions) and $N_\rho$ (the number of spectral densities to which a network is exposed during the training) are finite (see FIG.~\ref{fig:cartoon1});

\item the fact that at fixed 
\begin{flalign}\label{eq:N}
\vec{N}=(N_\mathrm{n},N_\mathrm{b},N_{\rho})
\end{flalign}
the answer of a network cannot be exact, and therefore has to be associated with an error, is taken into account by introducing an ensemble of machines, with $N_\mathrm{r}$ replicas, and by estimating this error by studying the distribution of the different $N_\mathrm{r}$ answers in the $N_\mathrm{r}\mapsto \infty$ limit (see FIG.~\ref{fig:cartoon2});

\item once the network (and statistical) errors at fixed $\vec N$ are given, we can study numerically the $\vec N\mapsto \infty$ limits and also quantify, reliably, the additional systematic errors associated with these unavoidable extrapolations  (see FIG.~\ref{fig:cartoon3}).

\end{itemize}

We are now going to provide the details concerning the choice of the functional basis that we used to parametrize the spectral densities and to build our training sets.

\subsection{The functional-basis} \label{sec:basis}
In our strategy we envisage studying numerically the limit $N_\mathrm{b}\mapsto \infty$ and, therefore, provided that the systematic errors associated with this extrapolation are properly taken into account, there is no reason to prefer a particular basis w.r.t.\ any other. For our numerical study we used the Chebyshev polynomials of the first kind as basis functions (see for example Ref.~\cite{boyd2001chebyshev}). 

The Chebyshev polynomials $T_n(x)$ are defined for $x \in [-1,1]$ and satisfy the orthogonality relations
\begin{flalign}\label{eq:OrthoCheby}
	\int_{-1}^{+1} \mathrm{d} x \frac{T_n(x)T_m(x)}{\sqrt{1-x^2}} =\begin{cases}
	0	&  n \neq m, \\ 
	\frac{\pi}{2}	& n =m\neq 0 \\ 
	\pi	&  n=m=0
	\end{cases} \;.
\end{flalign}
In order to use them as a basis for the spectral densities that live in the energy domain $E\in [E_0,\infty)$, we introduced the exponential map
\begin{flalign}
x(E)=1-2e^{-E}
\end{flalign}
and set
\begin{flalign}
B_n(E) = T_n(x(E))-T_n(x(E_0))\;.
\end{flalign}
Notice that the subtraction of the constant term $T_n(x(E_0))$ has been introduced in order to be able to cope with the fact that hadronic spectral densities vanish below a threshold energy $E_0\ge 0$ that we consider an unknown of the problem. With this choice, the unsmeared spectral densities that we use to build our training sets are written as
\begin{flalign}
\rho(E;N_\mathrm{b})= \theta(E-E_0)\sum_{n=0}^{N_\mathrm{b}}c_n 
\left[ T_n\left(x(E)\right)-T_n\left(x(E_0)\right) \right]\;,
\label{eq:rho_cheby}
\end{flalign}
and vanish identically for $E\le E_0$. Once $E_0$ and the coefficients $c_n$ that define $\rho(E;N_\mathrm{b})$ are given, the correlator and the smeared spectral density associated with $\rho(E;N_\mathrm{b})$ can be calculated by using  Eq.~(\ref{eq:inputC}) and Eq.~(\ref{eq:outputRho})\footnote{By representing also the smearing kernels and the basis functions on a Chebyshev basis, the orthogonality relations of Eq.~(\ref{eq:OrthoCheby}) can conveniently be exploited to speedup this step of the numerical calculations.}.

 \begin{figure}[t]	
 \includegraphics[width=\columnwidth]{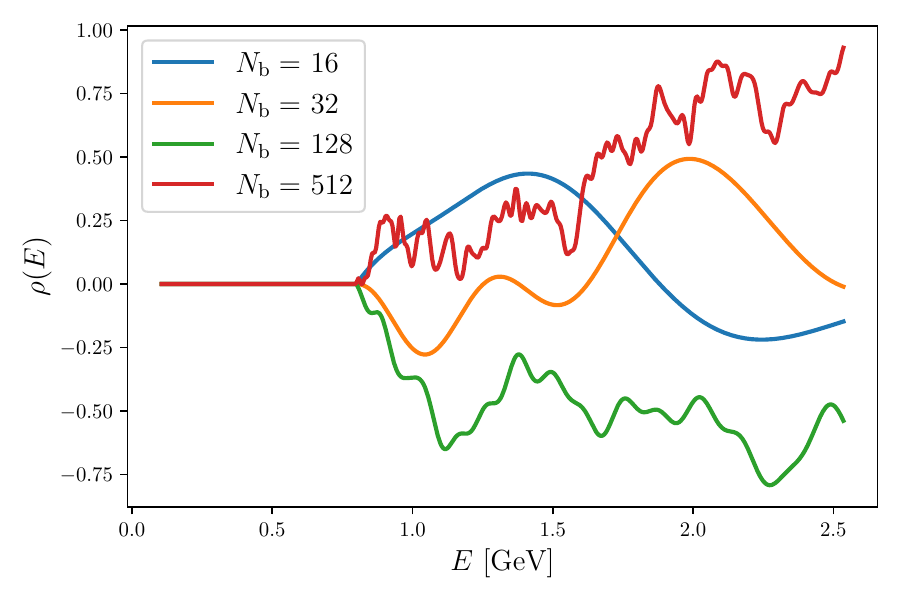}
\caption{\small Examples of unsmeared spectral densities generated according to Eq.~\ref{eq:rho_cheby} using the Chebyshev polynomial as basis functions for different $N_\mathrm{b}$. For all the examples we set $E_0=0.8$ GeV. As it can be seen, by generating the $c_n$ coefficients according to Eq.~(\ref{p10}), the larger the number of terms of the Chebyshev series the richer is the  behaviour of the spectral density in terms of local structures.}\label{fig:training_set1}
\end{figure}
\begin{figure}[t]
\includegraphics[width=\columnwidth]{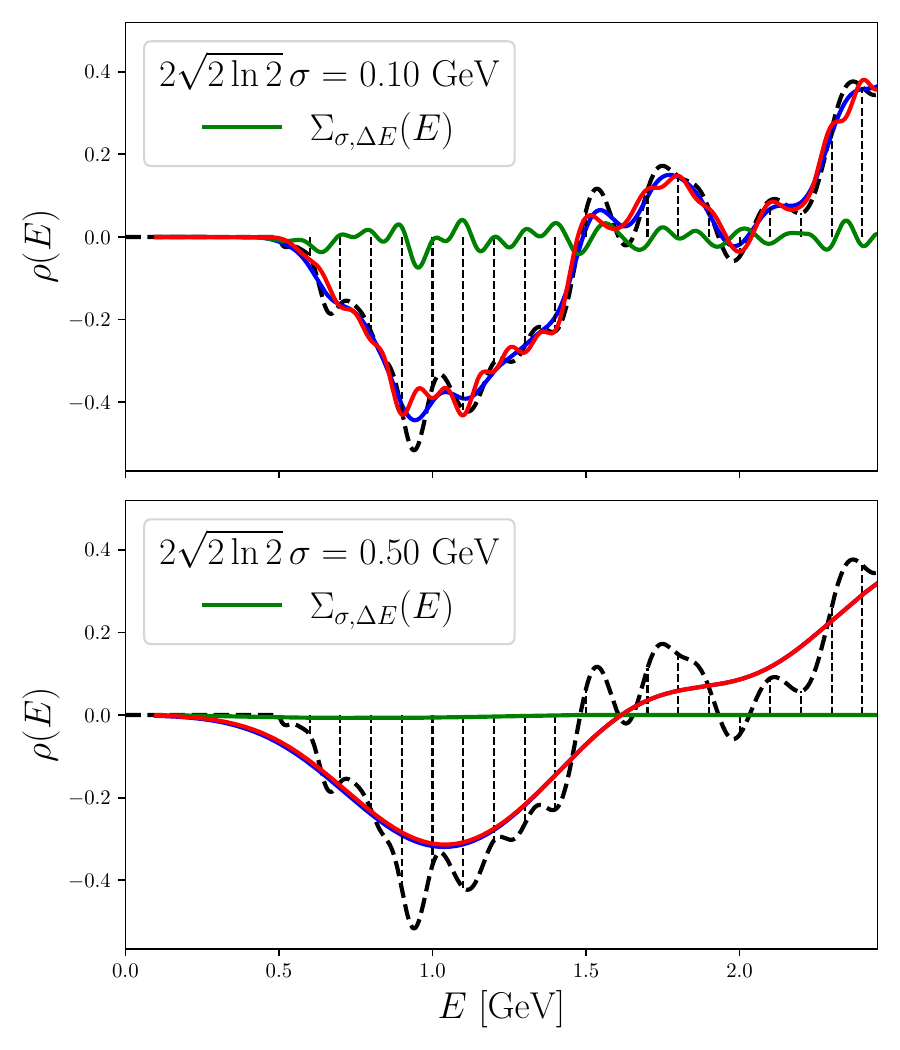}
\caption{\small The black dashed curve is a generic continuous unsmeared spectral density \(\rho(E)\)  while the vertical dashed lines indicate the energy sampling defining the support of $\rho_\delta(\omega)$ according to Eq.~(\ref{eq:fewlines2}).
 			The blue and red curves are the smeared version of respectively  $\rho(\omega)$ and $\rho_\delta(\omega)$. The green curve represents $\Sigma_{\sigma,\Delta E}(E)=\hat{\rho}_\sigma(E)-\hat{\rho}_{\delta,\sigma}(E)$.
 			The spacing is set to $\Delta E=0.1$ GeV.  The smearing kernel is a normalized Gaussian function of width $\sigma$ and as resolution width we refer to the full width at half maximum $2 \sqrt{2\ln 2}\, \sigma$ . 	 \emph{Top-panel}: the resolution width  is 	equal to $\Delta E$	 and it is such that the single peaks are resolved. 	\emph{Bottom-panel}: in this case $2\sqrt{2 \ln 2} \,\sigma \gg \Delta E$ and the two smeared functions are almost undistinguishable.
} 
\label{fig:PeaksVSContinuous}
\end{figure}

Each $\rho(E;N_\mathrm{b})$ that we used in order to populate our training sets has been obtained by choosing $E_0$ randomly in the interval $[0.2,1.3]$~GeV with a uniform distribution and by inserting in Eq.~(\ref{eq:rho_cheby}) the coefficients
\begin{flalign}\label{p10}
c_0 =r_0\;;
\qquad
\qquad
c_n = \frac{r_n}{n^{1+\varepsilon}}\;,\quad n>0\;,
\end{flalign}
where the $r_n$'s are $N_\mathrm{b}$ uniformly distributed random numbers in the interval $[-1,1]$ and $\varepsilon$ is a non-negative parameter that we set to $10^{-7}$. Notice that with this choice of the coefficients $c_n$ the Chebyshev series of Eq.~(\ref{eq:rho_cheby}) is convergent in the $N_\mathrm{b}\mapsto \infty$ limit and that the resulting spectral densities can be negative and in general change sign several times in the interval $E\in [0,\infty)$. Notice also that up to a normalization constant, that it will turn to be irrelevant in our training strategy in which the input data are standardized as explained in subsection~\ref{sec:preprocessing}, the choice of the interval $[-1,1]$ for the $r_n$ random numbers is general.

A few examples of unsmeared spectral densities generated according to Eq.~\ref{eq:rho_cheby} are shown in FIG.~\ref{fig:training_set1}. As it can be seen, with the choice of the coefficients $c_n$ of Eq.~(\ref{p10}), the larger is the number of terms in the Chebyshev series of Eq.~(\ref{eq:rho_cheby}) the richer is the behaviour of the resulting unsmeared spectral densities in terms of local structures. 

This is an important observation and a desired feature. Indeed, a natural concern about the choice of Chebyshev polynomials as basis functions is the fact that we are thus sampling the space of regular functions while, as we pointed out in section~\ref{sec:problem}, on a finite volume the lattice QCD spectral density is expected to be a discrete sum of Dirac-delta peaks (see Eq.~(\ref{eq:wilddistrib})). Here we are relying on the fact that the inputs of our networks will be Euclidean correlators, that in fact are smeared spectral densities, and will also be asked to provide as output the smeared spectral densities $\hat \rho_\sigma(E)$. This allows us to assume that, provided that the energy smearing parameter $\sigma$ and  $N_\mathrm{b}$ are sufficiently large, the networks will be exposed to sufficiently general training sets to be able to extract the correct smeared spectral densities within the quoted errors. To illustrate this point we consider in FIG.~\ref{fig:PeaksVSContinuous} a continuous spectral density $\rho(E)$ and approximate its smeared version through a Riemann's sum according to
\begin{flalign}\label{eq:fewlines1}
\hat{\rho}_\sigma(E)
&=\int_{E_0}^{\infty} \mathrm{d} \omega\, K_\sigma(E,\omega) \rho(\omega) 
\nonumber \\
&=  \Delta E\, \sum_{n=0}^\infty \,K_\sigma(E,\omega_n) \rho(\omega_n) + \Sigma_{\sigma,\Delta E}(E) 
\nonumber \\
&= \int_{E_0}^{\infty} \mathrm{d}\omega\, K_\sigma(E,\omega) 	\rho_\delta(\omega) +\Sigma_{\sigma,\Delta E}(E),
\end{flalign}
where
\begin{flalign} \label{eq:fewlines2}
\rho_\delta(\omega)  \equiv  
\Delta E\, \sum_{n=0}^\infty   \rho(\omega)\delta(\omega-\omega_n)\;,
\quad
\omega_n=E_0+n\Delta E\;.
\end{flalign}
The previous few lines of algebra show that the smearing of a continuous function $\rho(\omega)$ can be written as the smearing of the distribution defined in Eq.~(\ref{eq:fewlines2}), a prototype of the finite-volume distributions of Eq.~(\ref{eq:wilddistrib}), plus the approximation error $\Sigma_{\sigma,\Delta E}(E)=\hat{\rho}_\sigma(E)-\hat{\rho}_{\delta,\sigma}(E)$. The quantity $\Sigma_{\sigma,\Delta E}(E)$, depending upon the spacing $\Delta E$ of the Dirac-delta peaks and the smearing parameter $\sigma$, is expected to be sizeable when $\sigma\le \Delta E$ and to become irrelevant in the limit $\sigma\gg \Delta E$. A quantitative example is provided in FIG.~\ref{fig:PeaksVSContinuous} where $\Sigma_{\sigma,\Delta E}(E)$ is shown at fixed $\Delta E$ for two values of $\sigma$. In light of this observation, corroborated by the extensive numerical analysis that we have performed at the end of the training sessions to validate our method (see subsection~\ref{sec:velidation} and, in particular, FIG.~\ref{fig:res_peaks}), we consider justified the choice of Chebyshev polynomials as basis functions provided that, as is the case in this work, the energy smearing parameter $\sigma$ is chosen sufficiently large to not be able to resolve the discrete structure of the finite-volume spectrum.

\subsection{\label{sec:noise}
Building the training sets}

Having provided all the details concerning the functional basis that we use to parametrize the unsmeared spectral densities, we can now explain in details the procedure that we used to build our training sets.

A training set $\mathcal{T}_\sigma(N_\mathrm{b},N_\rho)$ contains $N_\rho$ input-output pairs. Each pair is obtained by starting from a random unsmeared spectral density, parametrized at fixed $N_\mathrm{b}$ according to Eq.~(\ref{eq:rho_cheby}) and generated as explained in the previous subsection. Given the unsmeared spectral density $\rho(E;N_\mathrm{b})$, we then compute the corresponding correlator vector $\vec C$ (by using Eq.~(\ref{eq:inputC})) and, for the two values of $\sigma$ that we used in this study, the smeared spectral densities vectors $\vec{\hat \rho_\sigma}$ (by using Eq.~(\ref{eq:outputRho})). From each pair $(\vec C,\vec{\hat \rho_\sigma})$ we then generate an element $(\vec{C}_\mathrm{noisy},\vec{\hat \rho_\sigma})$ of the training set at fixed $N_\mathrm{b}$ and $\sigma$,
\begin{flalign}
&
\rho^i(E;N_\mathrm{b})
\mapsto 
(\vec{C},\vec{\hat \rho_\sigma})^i
\mapsto 
(\vec{C}_\mathrm{noisy},\vec{\hat \rho_\sigma})^i
\in
\mathcal{T}_\sigma(N_\mathrm{b},N_\rho)\;,
\nonumber \\
\nonumber \\
&
i=1,\cdots, N_\rho\;,
\label{eq:addingnoise}
\end{flalign}
that we obtain, as we are now going to explain, by distorting the correlator $\vec C$ using the information provided by the noise of the lattice correlator $\vec{C}_\mathrm{latt}$ (see Section~\ref{sec:lattice_data}).

\begin{figure}
\includegraphics[width=\columnwidth]{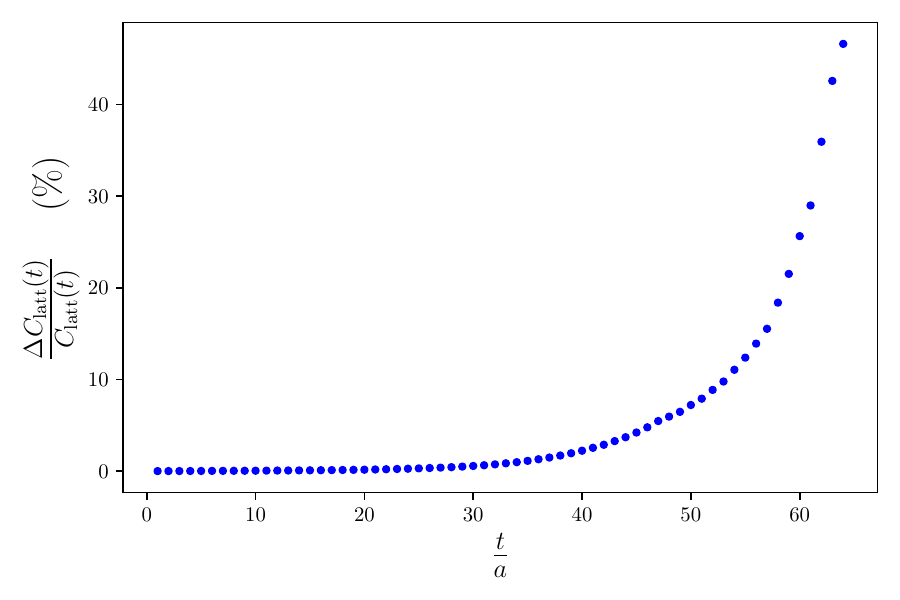}
\caption{\small Noise-to-signal ratio of the lattice correlator $C_\mathrm{latt}(t)$, discussed in Section~\ref{sec:lattice_data}, whose covariance matrix is used to inject the statistical noise in the training sets.}
\label{fig:SN_obs_B64}
\end{figure}

In order to cope with the presence of noise in the input data that have to be processed at the end of the training, it is extremely useful (if not necessary) to teach to the networks during the training to distinguish the physical content of the input data from noisy fluctuations. This is particularly important when dealing with lattice QCD correlators for which, as discussed in subsection~\ref{sec:inoutformat},  the noise-to-signal ratio grows exponentially for increasing Euclidean times (see FIG.~\ref{fig:SN_obs_B64}). A strategy to cope with this problem, commonly employed in the neural network literature, is to add Gaussian noise to the input data used in the training. There are several examples in the literature where neural networks are shown to be able to learn rather efficiently by employing this strategy of data corruption (see the already cited textbooks Refs.~\cite{books/lib/RasmussenW06,murphy2013machine}). 
According to our experience, it is crucially important that the structure of the noise used to distort the training input data resembles as much as possible that of the true input data. In fact, it is rather difficult to model the noise structure generated in Monte Carlo simulations and, in particular, the off-diagonal elements of the covariance matrices of lattice correlators. In the light of these observations we decided to use the covariance matrix $\hat{\Sigma}_\mathrm{latt}$ of the lattice correlator $\vec{C}_\mathrm{latt}$ that we are going to analyse in Section~\ref{sec:lattice_data} to obtain $\vec{C}_\mathrm{noisy}$ from $\vec C$. More precisely, given a correlator $\vec{C}$, we generate $\vec{C}_\mathrm{noisy}$ by extracting a random correlator vector from the multivariate Gaussian distribution
\begin{flalign}\label{eq:Multivariate}
	\mathbb{G}\left[\vec{C},\left(\frac{C(a)}{C_\mathrm{latt}(a)}\right)^2 \hat{\Sigma}_\mathrm{latt}\right]
\end{flalign}
having $\vec{C}$ as mean vector and as covariance the matrix $\hat{\Sigma}_\mathrm{latt}$ normalized by 
the factor $\left(\frac{C(a)}{C_\mathrm{latt}(a)}\right)^2$. 

In order to be able to perform a numerical study of the limits $\vec{N} \mapsto \infty$, we have generated with the procedure just described several training sets, corresponding to $N_\mathrm{b}=\{16,32,128,512\}$ and $N_\rho=\{50,100,200,400,800\}\times 10^3$. At fixed $N_\mathrm{b}$, each training set includes the smaller ones, i.e. the training set $\mathcal{T}_\sigma(N_\mathrm{b},100\times 10^3)$ includes the training set $\mathcal{T}_\sigma(N_\mathrm{b}, 50\times 10^3)$ enlarged with $50\times 10^3$ new samples.

\subsection{Data pre-processing}\label{sec:preprocessing}

A major impact in the machine-learning performance is played by the way the data are presented to the neural network. The learning turns to be more difficult when the input variables have different scales and this is exactly the case of Euclidean lattice correlators since the components of the input vectors decrease exponentially fast. The risk in this case is that the components small in magnitude are given less importance during the training or that some of the weights of the network become very large in magnitude, thus generating instabilities in the training process.  We found that a considerable improvement in the validation loss minimization is obtained by implementing the standardization of the input correlators (see next subsection and the central panel of FIG.~\ref{fig:losses_test}) and therefore used this procedure. 

The standardization procedure of the input consists in rescaling the data in such a way that all the components of the input correlator vectors in the training set are distributed with average 0 and variance 1.  For a given training set $\mathcal{T}_\sigma(N_\mathrm{b},N_\rho)$ we calculate the $N_T$-component vectors $\vec{\mu}$ and $\vec\gamma$ whose components are 
\begin{flalign}
	\mu(\tau)=\frac{1}{N_\rho}\sum_{i=1}^{N_\rho}C_i(a\tau)
\end{flalign}
and
\begin{flalign}
	\gamma(\tau)= \sqrt{\frac{\sum_{i=1}^{N_\rho} \left(C_i(a\tau)-\mu(\tau)\right)^2}{N_\rho}}	.
\end{flalign}
Each correlator in the training is then replaced by
\begin{flalign}\label{eq:preprocess}
	C_\mathrm{noisy}(a\tau)\mapsto C'_\mathrm{noisy}(a\tau)=\frac{C_\mathrm{noisy}(a\tau)-\mu(\tau)}{\gamma(\tau)}.
\end{flalign}
where $\vec{C}_\mathrm{noisy}$ is the distorted version of $\vec C$ discussed in the previous subsection. Notice that the vectors $\vec{\mu}$ and $\vec{\gamma}$ are determined from the training set before including the noise. Although the pre-processed correlators look quite different from the original ones, since the components are no longer exponential decaying, the statistical correlation in time is preserved by the standardization procedure.  

At the end of the training, the correlators fed into the neural network for validation or prediction have also to be pre-processed by using the same vectors $\vec{\mu}$ and $\vec{\gamma}$ used in the training.

\subsection{\label{sec:Training}Training an ensemble of machines}

Given a machine with $N_\mathrm{n}$ neurons we train it over the training set $\mathcal{T}_\sigma(N_\mathrm{b},N_\rho)$ 
by using as loss function the Mean Absolute Error (MAE)
\begin{flalign}\label{eq:loss}
\ell(\vec{w}) = \frac{1}{N_\rho}
\sum_{i=1}^{N_\rho}\left\vert\vec{\hat{\rho}}_{\sigma}^{\mathrm{pred},i}(\vec{w})-\vec{\hat{\rho}}_{\sigma}^i\right\vert\;,
\end{flalign}
where  $\vec{\hat{\rho}}_\sigma^{\mathrm{pred},i}(\vec{w})$ is the output of the neural network in correspondence of the input correlator $\vec{C}_\mathrm{noisy}^i$. In Eq.~(\ref{eq:loss}) we used the norm  $|\bm{\hat{\rho}}|=\sum_{j=1}^{N_E=47}|\bm{\hat{\rho}}_j|$ and we have explicitly shown the dependence of the predicted spectral density $\vec{\hat{\rho}}_{\sigma}^{\mathrm{pred},i}(\vec{w})$ upon the weights $\vec{w}$ of the network. 

At the beginning of each training session each weight $w_n$ (with $n=1,\cdots,N_\mathrm{n}$) is extracted from a Gaussian distribution with zero mean value and variance $0.05$. To end the training procedure we rely on the early stopping criterion: the training set is split into two subsets containing respectively the 80\% and 20\% of the entries of the training set $\mathcal{T}_\sigma(N_\mathrm{b},N_\rho)$. The larger subset is used to update the weights of the neural network with the gradient descent algorithm. The smaller subset is the so-called \textit{validation set} and we use it to monitor the training process. At the end of each epoch we calculate the loss function of Eq.~(\ref{eq:loss}) for the validation set, the so-called validation loss, and stop the training when the drop in the validation loss is less than $10^{-5}$ for 15 consecutive epochs. In the trainings performed in our analysis this occurs typically between epoch 150 and 200. The early stopping criterion provides an automatic way to prevent the neural network from overfitting the input data.

\begin{figure}[t!]
	\includegraphics[width=\columnwidth]{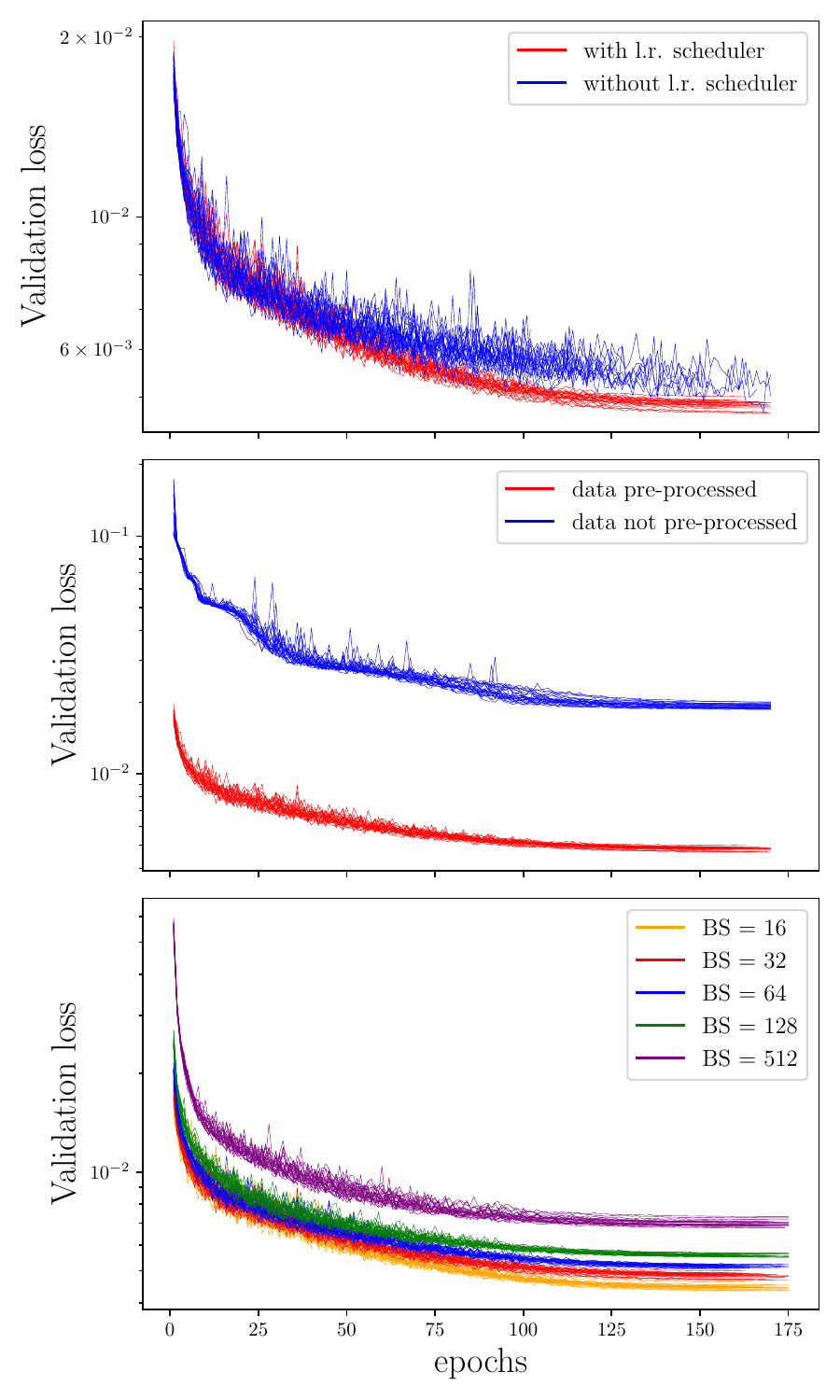}
	\caption{\small Numerical tests to optimize the training performances. The validation loss functions of the trainings of the $N_\mathrm{r}=20$ replica machines belonging to the same ensemble are plotted with the same color. All the trainings refer to \textit{arcL}, $N_\rho=50\times 10^3$, $N_\mathrm{b}=512$ and $\sigma=0.44$ GeV. If not otherwise specified the data are pre-processed, BS=32 and the learning rate scheduler is implemented. 
	\emph{Top panel}: Comparison of the validation losses with and without the learning scheduler rate defined in Eq.~(\ref{eq:lrsched}). 
	\emph{Central panel}:
		Comparison of the validation losses by implementing or not the input data pre-processing procedure described in subsection~\ref{sec:preprocessing}.
	\emph{Bottom panel}: Comparison of the validations losses at different values of the BS parameter.}
	\label{fig:losses_test}
\end{figure}

\begin{figure}[t!]
	\centering
	\includegraphics[width=\columnwidth]{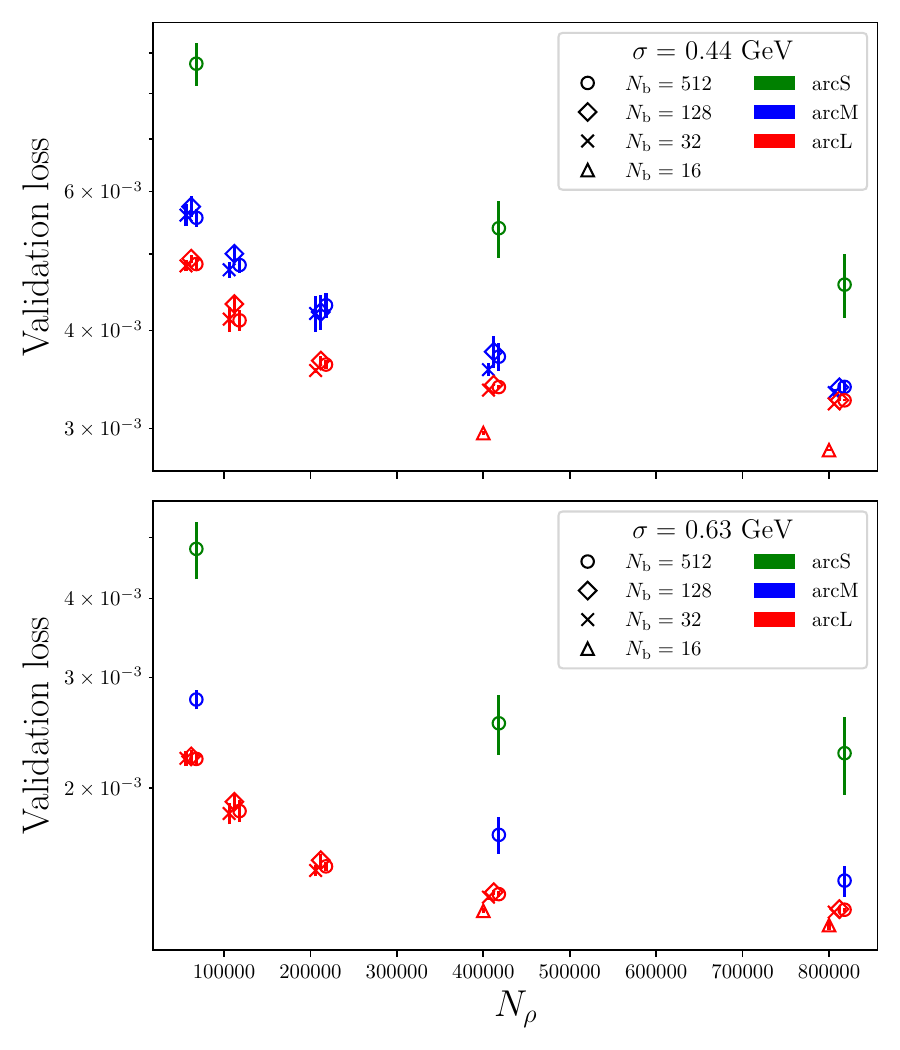}
	\caption{\small Validation losses at the end of the trainings as functions of the training set dimension $N_\rho$. The top-plot corresponds to $\sigma=0.44$~GeV and the bottom-plot to $\sigma=0.63$~GeV. The different colors correspond to the different architectures and different markers to the different number of basis functions $N_\mathrm{b}$ considered in this study. The error on each point correspond to the standard deviation of the distribution of the validation loss in an ensemble of $N_\mathrm{r}=20$ machines. As expected, the validation loss decreases by increasing $N_\rho$ (more general training) and/or $N_\mathrm{n}$ (larger and therefore smarter networks). Moreover, at fixed $N_\mathrm{n}$ and $N_\rho$ the neural network performs better at smaller values of $N_\mathrm{b}$ as consequence of the fact that the training set exhibits less complex features and learning them is easier.}\label{fig:final_losses}
\end{figure}

We implement a Mini-Batch Gradient Descent algorithm, with Batch Size (BS) set to 32, by using the  Adam optimizer \cite{kingma2017adam} combined with a learning rate decaying according to
\begin{flalign}\label{eq:lrsched}
	\eta(e)=\frac{\theta(e-25)\eta(e-1)}{1+e\cdot 4\times 10^{-4}},
\end{flalign}
where $e\in \mathbb{N}$ is the epoch and $\eta(0)=2\times 10^{-4}$ is the  starting value. The step-function is included so that the learning rate is unchanged during the first 25 epochs. 
Although a learning rate scheduler is not strictly mandatory, since the Adam optimizer already includes adaptive learning rates, we found that it provides an improvement in the convergence with less noisy fluctuations in the validation loss (see the top panel of FIG.~\ref{fig:losses_test}). 
Concerning the BS, we tested the neural network performance by starting from BS=512 and by halving it up to BS=16 (see bottom panel of FIG.~\ref{fig:final_losses}). Although the performance improves as BS decreases we set BS$=32$ in order to cope with the unavoidable slowing down of the training for smaller values of BS.

As we already stressed several times, at fixed $\vec N$ the answer of a neural network cannot be exact. In order to be able to study numerically the $\vec N\mapsto \infty$ limits, the error associated with the limited abilities of the networks at finite $\vec N$ \emph{has} to be quantified. To do this we introduce the \textit{ensemble of machines} by considering at fixed $\vec N$
\begin{flalign}
N_\mathrm{r}=20
\end{flalign}
machines with the same architecture and trained by using the same strategy. More precisely, each machine of the ensemble is trained by using a training set $\mathcal{T}_\sigma(N_\mathrm{b},N_\rho)$ obtained by starting from the same unsmeared spectral densities, and therefore from the same pairs $(\vec{C},\vec{\hat \rho_\sigma})$ (see Eq.~(\ref{eq:addingnoise})), but with different noisy input correlator vectors $\vec{C}_\mathrm{noisy}$. 

In FIG.~\ref{fig:final_losses} we show the validation loss as a function of $N_\rho$ for the different values of $N_\mathrm{b}$, the three different architectures and the two values of $\sigma$ considered in this study. Each point in the figure has been obtained by studying the distribution of the validation loss at the end of the training within the corresponding ensemble of machines and by using  respectively the mean and the standard deviation of each distribution as central value and error. The figure provides numerical evidences concerning the facts that
\begin{itemize}
\item networks with a finite number $N_\mathrm{n}$ of neurons, initialized with different weights and exposed to training sets containing a finite amount of information, provide different answers and this is a source of errors that have to be quantified;

\item the validation loss decreases for larger $N_\mathrm{n}$ because larger networks are able to assimilate a larger amount of information;

\item  the validation loss decreases for larger $N_\rho$ since, if $N_\mathrm{n}$ is sufficiently large, the networks learn more from larger and more general training sets;

\item at fixed $N_\mathrm{n}$ and $N_\rho$ the networks perform better at smaller values of $N_\mathrm{b}$ because the training sets exhibit less complex features and learning them is easier.
\end{itemize}

In order to populate the plots in FIG.~\ref{fig:final_losses} we considered several values of $N_\rho$ and $N_\mathrm{b}$ and this is rather demanding from the computational point of view. The training sets that we found to be strictly necessary to quote our final results are listed in TABLE~\ref{tab:refN}.  

\section{\label{sec:quote_results}
Estimating the total error and Validation Tests}

Having explained in the previous section the procedure that we used to train our ensembles of machines, we can now explain in details the procedure that we use to obtain our results. We start by discussing the procedure that we use to estimate the total error, taking into account both statistical and systematics uncertainties, and then illustrate the validation tests that we have performed in order to assess the reliability of this procedure.

\subsection{\label{sec:quotation}Procedure to estimate the total error}

The procedure that we use to quote our results for smeared spectral densities by using the trained ensembles of machines, illustrated in FIG.~\ref{fig:cartoon3}, is the following  
\begin{itemize}

\item given an ensemble of $N_\mathrm{r}$ machines trained at fixed $\vec N$, we feed in each machine $r$ belonging to the ensemble all the different bootstrap samples (or jackknife bins) $C_c(a\tau)$ of the input correlator ($c=1,\cdots, N_\mathrm{c}$)  and obtain a collection of results $\hat{\rho}_\sigma^\mathrm{pred}(E,\vec{N},c,r)$ for the smeared spectral density that depend upon $\vec N$, $c$ and $r$; 

\item at fixed $\vec N$ and $r$ we compute the bootstrap (or jackknife) central values $\hat{\rho}_\sigma^\mathrm{pred}(E,\vec{N},r)$ and statistical errors $\Delta_\sigma^\mathrm{latt}(E,\vec{N},r)$ and average them over the ensemble of machines,
\begin{flalign}
&
\hat{\rho}_\sigma^\mathrm{pred}(E,\vec{N})=\frac{1}{N_\mathrm{r}}\sum_{r=1}^{N_\mathrm{r}}\hat{\rho}_\sigma^\mathrm{pred}(E,\vec{N},r)\;,
\nonumber \\
&
\Delta_\sigma^\mathrm{latt}(E,\vec{N})=\frac{1}{N_\mathrm{r}}\sum_{r=1}^{N_\mathrm{r}}\Delta_\sigma^\mathrm{latt}(E,\vec{N},r)\;;
\end{flalign}

\item by computing the standard deviation over the ensemble of machines of $\hat{\rho}_\sigma^\mathrm{pred}(E,\vec{N},r)$, we obtain an estimate of the error, that we call $\Delta_\sigma^\mathrm{net}(E,\vec{N})$, associated with the fact that at fixed $\vec N$ the answer of a network cannot be exact;

\item both $\Delta_\sigma^\mathrm{latt}(E,\vec{N})$ and $\Delta_\sigma^\mathrm{net}(E,\vec{N})$ have a statistical origin. The former, $\Delta_\sigma^\mathrm{latt}(E,\vec{N})$, comes from the limited statistics of the lattice Monte Carlo simulation while the latter, $\Delta_\sigma^\mathrm{net}(E,\vec{N})$, comes from the statistical procedure that we used to populate our training sets $\mathcal{T}_\sigma(N_\mathrm{b},N_\rho)$ and to train our ensembles of machines at fixed $N_\mathrm{b}$. We sum them in quadrature and obtain an estimate of the statistical error at fixed $\vec N$,
\begin{flalign}\label{eq:stat}
	\Delta_\sigma^\mathrm{stat}(E,\vec{N}) = \sqrt{\left[\Delta_\sigma^\mathrm{latt}(E,\vec{N})\right]^2+\left[\Delta_\sigma^\mathrm{net}(E,\vec{N})\right]^2}\;;
\end{flalign}

\item we then study numerically the $\vec N\mapsto \infty$ limits by using a data-driven procedure. We quote as central value and statistical error of our final result 
\begin{flalign}\label{eq:final_pred}
&
\hat{\rho}_\sigma^\mathrm{pred}(E) \equiv \rho_\sigma^\mathrm{pred}(E,\vec{N}^\mathrm{max})\;,
\nonumber \\
\nonumber \\
&
\Delta_\sigma^\mathrm{stat}(E) \equiv \Delta_\sigma(E,\vec{N}^\mathrm{max})\;,
\end{flalign} 
where $\vec{N}^\mathrm{max}$, given in Table~\ref{tab:refN}, is the vector with the largest components among the vectors $\vec N$ considered in this study;

\item in order to check the numerical convergence of the $\vec N\mapsto \infty$ limits and to estimate the associated systematic uncertainties $\Delta_\sigma^{X}(E)$, where 
\begin{flalign}
X=\{\rho,\mathrm{n},\mathrm{b}\}\;,
\end{flalign}
we define the reference vectors $\vec{N}_X^\mathrm{ref}$ listed in Table~\ref{tab:refN}. We then define the pull variables
\begin{flalign}
P_\sigma^{X}(E) = \frac{\left|\hat{\rho}_\sigma^{\mathrm{pred}}(E)-\hat{\rho}_\sigma^{\mathrm{pred}}(E,\vec{N}^\mathrm{ref}_X)\cdot\right|}{\sqrt{\big[\Delta_\sigma^\mathrm{stat}(E)\big]^2+\big[\Delta_\sigma^\mathrm{stat}(E,\vec{N}^\mathrm{ref}_X)\big]^2}}\;,
\end{flalign}
and then we weight the absolute value of the difference $\left|\hat{\rho}_\sigma^{\mathrm{pred}}(E)-\hat{\rho}_\sigma^{\mathrm{pred}}(E,\vec{N}^\mathrm{ref}_X)\right|$ with the Gaussian probability that this is not due to a statistical fluctuation, 
\begin{flalign}\label{eq:DeltaXsigma}
\Delta^{X}_\sigma(E) 
= \left\vert\hat{\rho}_\sigma^{\mathrm{pred}}(E)-\hat{\rho}_\sigma^{\mathrm{pred}}(E,\vec{N}^\mathrm{ref}_X)\right\vert \mathrm{erf} \left(\frac{|P_\sigma^{X}(E)|}{\sqrt{2}}\right);
\end{flalign}

\item the total error that we associate to our final result $\hat{\rho}_\sigma^\mathrm{pred}(E)$ is finally obtained according to
\begin{flalign}\label{eq:tot_error}
&
\Delta_\sigma^\mathrm{tot}(E) 
\nonumber \\
&
= \sqrt{
\left[\Delta_\sigma^\mathrm{stat}(E)\right]^2+
\left[\Delta_\sigma^\rho(E)\right]^2+
\left[\Delta_\sigma^\mathrm{b}(E)\right]^2+
\left[\Delta_\sigma^\mathrm{n}(E)\right]^2}\;.
\end{flalign}%
\end{itemize}
%

%
\begin{table}[t!]
	\begin{ruledtabular}
		\begin{tabular}{cccc}
			Label & $N_\mathrm{n}$ & $N_\mathrm{b}$ & $N_\rho$  \\  \hline
			$\vec{N}^\mathrm{max}$            & arcL & 512 & $800\times 10^3$ \\ 	\hline 	
			$\vec{N}_\mathrm{b}^\mathrm{ref}$   & arcL & 16  & $800\times 10^3$  \\ 
			$\vec{N}_\mathrm{n}^\mathrm{ref}$   & arcS & 512 &	$800\times 10^3$  \\ 
			$\vec{N}_\rho^\mathrm{ref}$ 		 & arcL & 512 & $50\times 10^3$ \\ 		 		
		\end{tabular}
	\end{ruledtabular}
	\caption{\small List of the vectors $\vec N$ used to estimate the systematic errors associated with the $\vec{N}\mapsto \infty$ limits. With these choices the errors $\Delta_\sigma^X(E)$ defined in Eq.~(\ref{eq:DeltaXsigma}) are maximised on our setup and, consequently, Eq.~(\ref{eq:tot_error}) provides a conservative estimate of the total error.}
	\label{tab:refN} 
\end{table}
Some remarks are in order here. We don't have a theoretical understanding neither of the dependence of the statistical errors $\Delta_\sigma^\mathrm{latt}(E,\vec{N})$ and $\Delta_\sigma^\mathrm{net}(E,\vec{N})$ upon $E$ and $\vec{N}$ (see Appendix~\ref{sec:netVSlatt} for a numerical investigation) nor of the rates of convergence of the $\vec N\mapsto \infty$ limits. Gaining this theoretical understanding is a task that goes far beyond the scope of this paper. The procedure that we have devised to quote our results, that at first sight might appear too complicated, has therefore to be viewed as just one of the possible ways to perform conservative plateaux-analyses of the $\vec N\mapsto \infty$ limits. This explains our choice of the points $\vec{N}^\mathrm{ref}_X$ given in Table~\ref{tab:refN} that, in our numerical setup, provide the most conservative estimates of the systematic errors. The stringent validation tests that we have performed with mock data, and that we are going to discuss below, provide a quantitative evidence that the results obtained by implementing this procedure can in fact be trustworthy used in phenomenological applications.

\subsection{Validation}\label{sec:velidation}
\begin{figure}[t!]
	\includegraphics[width=\columnwidth]{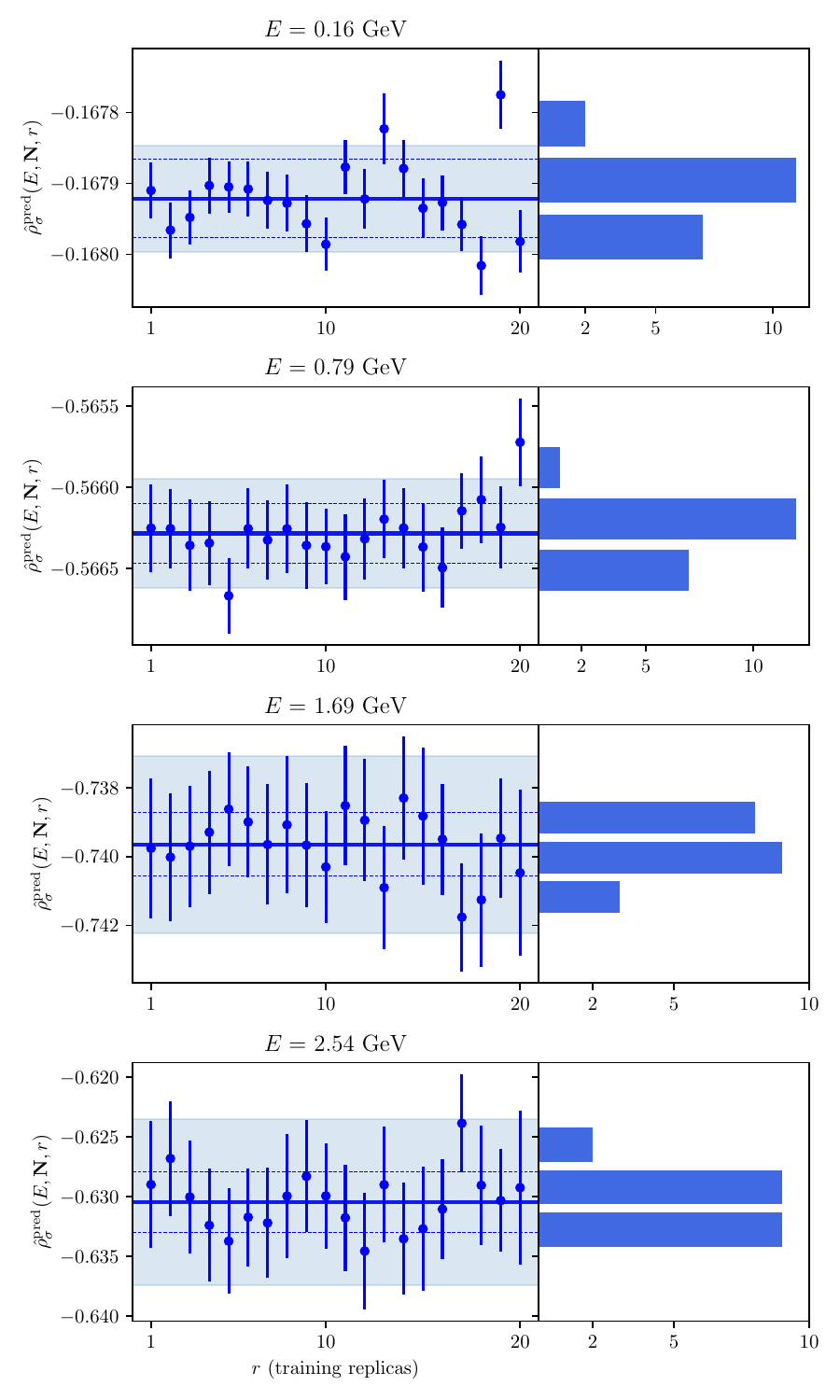}
	\caption{\small Each row corresponds to a different energy, as written on the top of each plot, and in all cases we set $\sigma=0.44$~GeV and $\vec{N}=\vec{N}^\mathrm{max}=(\mathrm{arcL},512,800\times 10^3)$. \emph{Left panels}: the $x$-axes corresponds to the replica index $r=1,\cdots,N_\mathrm{r}=20$ running over the entries of the ensemble of machines. The plotted points correspond to the results $\hat{\rho}_\sigma^\mathrm{pred}(E,\vec{N},r) \pm \Delta_\sigma^\mathrm{latt}(E,\vec{N},r)$ obtained by computing the bootstrap averages and errors of the $c=1,\cdots,N_\mathrm{c}=800$ results $\hat{\rho}_\sigma^\mathrm{pred}(E,\vec{N},c,r)$. \emph{Right panels}: distributions of the central values $\hat{\rho}_\sigma^\mathrm{pred}(E,\vec{N},r)$. The means of these distributions correspond to the results $\hat{\rho}_\sigma^\mathrm{pred}(E,\vec{N})$ that are shown in the left panels as solid blue lines. The widths correspond to the network errors $\Delta_\sigma^\mathrm{net}(E,\vec{N})$ that are represented in the left panels with two dashed blue lines. The blue bands in the left panels correspond to $\Delta_\sigma^\mathrm{stat}(E,\vec{N})$, defined in Eq.~(\ref{eq:stat}).}
	\label{fig:example1_stat_s4.2}
\end{figure}
\begin{figure}[t!]
	\includegraphics[width=\columnwidth]{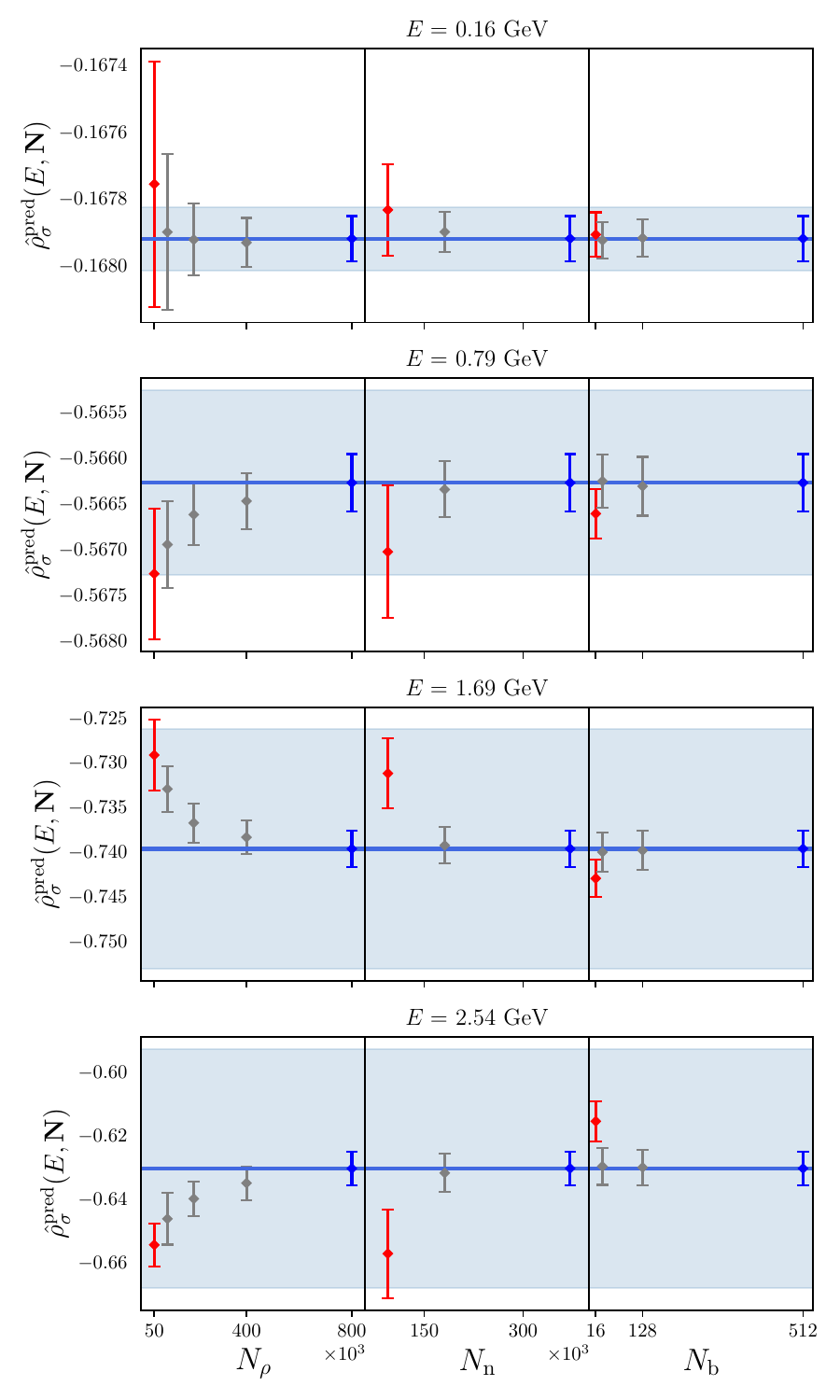}
	\caption{\small 
		Numerical study of the $\vec{N}\mapsto \infty$ limits. Each row corresponds to a different energy, as written on the top of each plot, and all cases we set $\sigma=0.44$ GeV. 
		\emph{Left panels:} Study of the limit $N_\rho \mapsto \infty$  at fixed $N_\mathrm{b}=512$ and $N_\mathrm{n}=$arcL. 
		\emph{Central panels}: Study of the  limit $N_\mathrm{n} \mapsto \infty$  at fixed $N_\mathrm{b}=512$ and $N_\rho=800\times 10^3$. 
		\emph{Right panels}: Study of the limit $N_\mathrm{b} \mapsto \infty$  at fixed $N_\mathrm{n}=$arcL and $N_\rho=800\times 10^3$. 
		\emph{All panels}: The horizontal blue line is the final central value, corresponding to $\vec{N}=\vec{N}^\mathrm{max}=(\mathrm{arcL},512,800\times 10^3)$ (blue points) and it is common to all the panels within the same row. The blue band represents instead  $\Delta_\sigma^\mathrm{tot}(E)$ calculated by the combination in quadrature of all the errors (see Eq.~\ref{eq:tot_error}). The red points correspond to $\vec{N}_X^\mathrm{ref}$ and the grey ones to the other trainings, see TABLE~\ref{tab:refN}.}\label{fig:example1_limits_s4.2}
\end{figure}
\begin{figure}[t!]
	\includegraphics[width=\columnwidth]{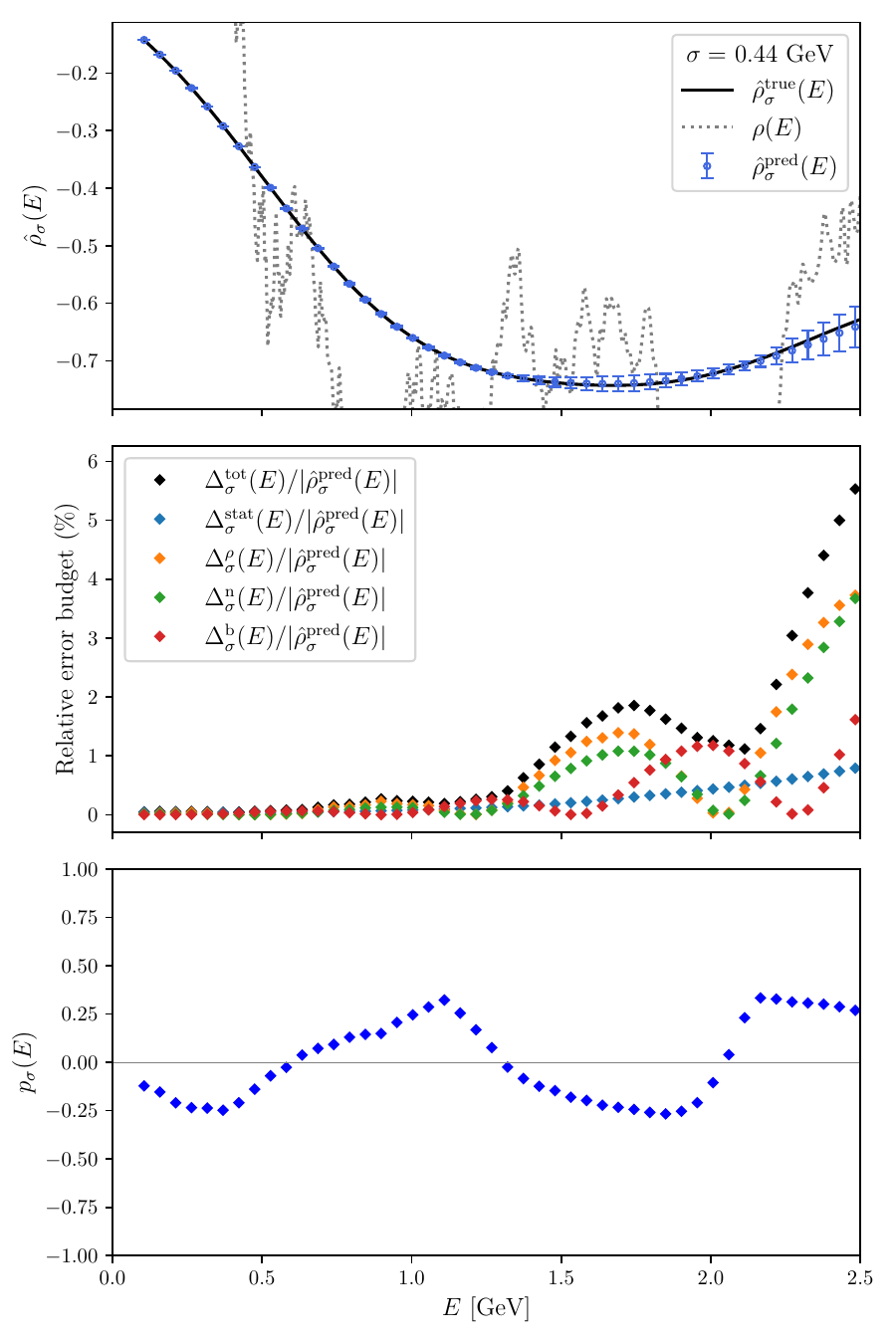}
	\caption{\small Final results for $\sigma=0.44$ GeV. \emph{Top panel}: The final results predicted by the neural networks (blue points) is compared with the true result (solid black). The dotted grey curve in the background represents the unsmeared spectral density. \emph{Central panel}: Relative error budget as a function of the energy. \emph{Bottom panel}: Deviation of the predicted result from the true one in units of the standard deviation, see Eq.~(\ref{eq:signficance}).}
	\label{fig:example1_final_s4.2}
\end{figure}
%
%
%
In order to illustrate how the method described in the previous subsection can be applied in practice, we now consider an unsmeared spectral density $\rho(E)$ that has not been used in any of the trainings and that has been obtained as described in subsection~\ref{sec:basis}, i.e.\ by starting from Eq.~(\ref{eq:rho_cheby}), but this time with $N_\mathrm{b}=1024$, i.e.\ a number of basis functions which is twice as large as the largest $N_\mathrm{b}$ employed in the training sessions. 

From this $\rho(E)$ we then calculate the associated correlator $\vec{C}$ and the smeared spectral density corresponding to $\sigma=0.44$ GeV. We refer to the true smeared spectral density as $\hat{\rho}_\sigma^\mathrm{true}(E)$, while we call $\hat{\rho}_\sigma^\mathrm{pred}(E)$ the final predicted result. 
Both the unsmeared spectral density $\rho(E)$ (grey dotted curve, partially visible) and the expected exact result $\hat{\rho}^\mathrm{true}_\sigma(E)$ (solid black curve) are shown in the top panel of FIG.~\ref{fig:example1_final_s4.2}. 

Starting from the exact correlator $\vec{C}$ corresponding to $\rho(E)$ we have then generated $N_\mathrm{c}=800$ bootstrap samples from the distribution of Eq.~(\ref{eq:Multivariate}), thus simulating the outcome of a lattice Monte Carlo simulation. The $N_\mathrm{c}$ samples have been fed into each trained neural network and we collected the answers $\hat{\rho}_\sigma^\mathrm{pred}(E,\vec{N},c,r)$ that, as shown in FIG.~\ref{fig:example1_stat_s4.2} for a selection of the considered values of $E$, we have then analysed to obtain $\Delta_\sigma^\mathrm{latt}(E,\vec{N})$ and $\Delta_\sigma^\mathrm{net}(E,\vec{N})$.

The next step is now the numerical study of the limits $\vec{N}\mapsto \infty$ that we illustrate in FIG.~\ref{fig:example1_limits_s4.2} for the same values of $E$ considered in FIG.~\ref{fig:example1_stat_s4.2}. The limit $N_\rho \mapsto \infty$ (left panels) is done at fixed $N_\mathrm{b}=512$ and $N_\mathrm{n}=$arcL. The limit $N_\mathrm{n} \mapsto \infty$ (central panels) is done at fixed $N_\mathrm{b}=512$ and $N_\rho=800\times 10^3$. The limit $N_\mathrm{b} \mapsto \infty$ (right panels) is done at fixed $N_\mathrm{n}=$arcL and $N_\rho=800\times 10^3$. As it can be seen, the blue points, corresponding to our results $\hat{\rho}_\sigma^\mathrm{pred}(E) \pm \Delta_\sigma^\mathrm{stat}(E)$, are always statistically compatible with the first grey point on the left of the blue one in each plot. The red points correspond to the results $\hat{\rho}_\sigma^\mathrm{pred}(E,\vec{N}_X^\mathrm{ref}) \pm \Delta_\sigma^\mathrm{stat}(E,\vec{N}_X^\mathrm{ref})$ that we use to estimate the systematic uncertainties $\Delta_\sigma^X(E)$. The blue band correspond to our estimate of the total error $\Delta_\sigma^\mathrm{tot}(E)$. 

In the top panel of FIG.~\ref{fig:example1_final_s4.2} we show the comparison of our final results $\hat{\rho}_\sigma^\mathrm{pred}(E) \pm \Delta_\sigma^\mathrm{tot}(E)$ (blue points) with the true smeared spectral density $\hat{\rho}_\sigma^\mathrm{true}(E)$ (black curve) that in this case is exactly known. The central panel in FIG.~\ref{fig:example1_final_s4.2} shows the relative error budget as a function of the energy. As it can be seen, the systematics errors represent a sizeable and important fraction of the total error, particularly at large energies. The bottom panel of FIG.~\ref{fig:example1_final_s4.2} shows the pull variable
\begin{flalign}\label{eq:signficance}
	p_\sigma(E)=\frac{\hat{\rho}_\sigma^\mathrm{pred}(E)-\hat{\rho}_\sigma^\mathrm{true}(E)}{\Delta_\sigma^\mathrm{tot}(E)}
\end{flalign}
and, as it can be seen, by using the proposed procedure to estimate the final results and their errors, no significant deviations from the true result have been observed in this particular case.

\begin{figure}[t!]
	\includegraphics[width=\columnwidth]{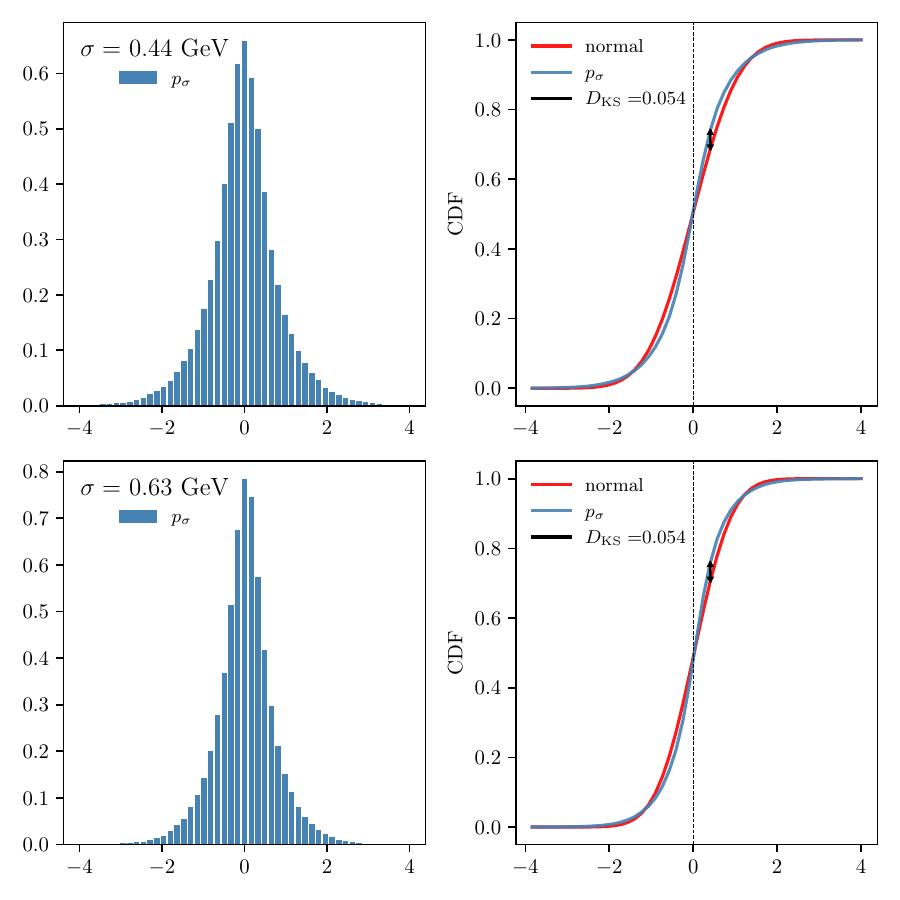}
	\caption{\small \emph{Left panel}: Normalized  distribution of the significance defined in Eq.~(\ref{eq:signficance}). The distribution is calculated over 2000 validation samples generated on the Chebyshev functional space. 
		\emph{Right panel}: comparison of the normalized cumulative distribution functions (CDF) of the observed distribution of $p_\sigma$ (blue) and of the normal one obtained from the mean and variance of $p_\sigma$ (red). The black arrow represents the Kolmogorov–Smirnov statistics ($D_\mathrm{KS}$), i.e. the magnitude and the position of the maximum deviation between the two CDFs in absolute value. }
\label{fig:cheby_sig}
\end{figure}
\begin{figure}[h!]
	\includegraphics[width=\columnwidth]{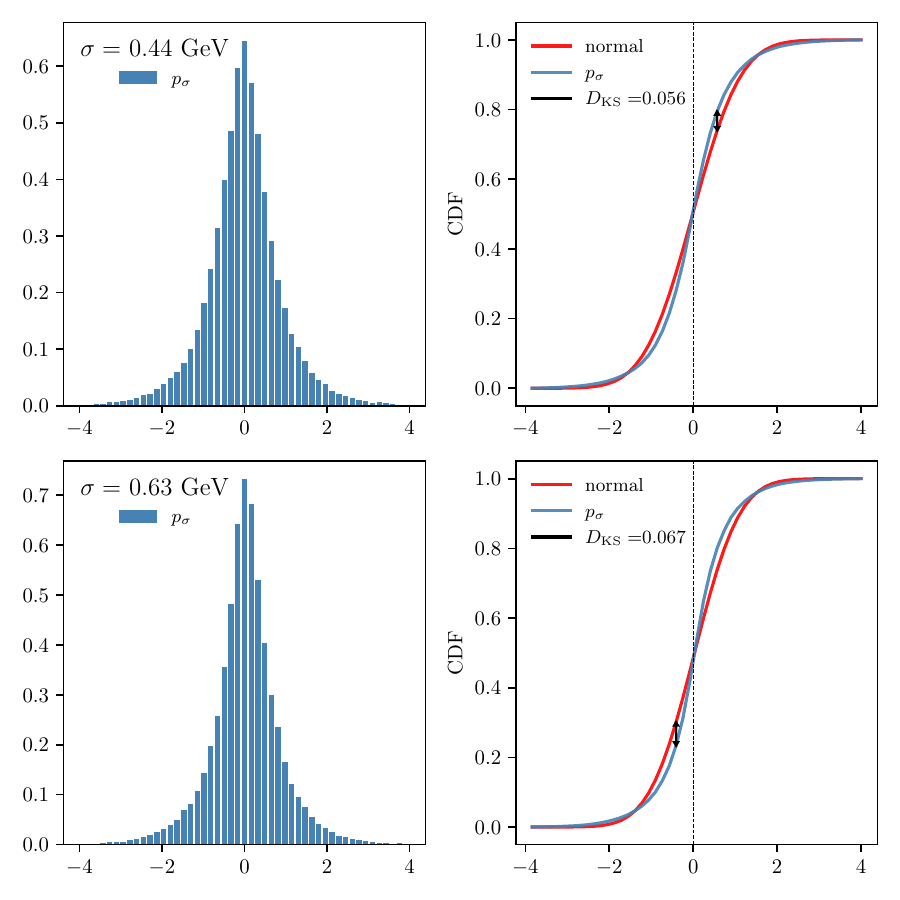}
	\caption{\small The same as FIG.~\ref{fig:cheby_sig} but for 2000 samples of unsmeared spectral densities generated according to Eq.~(\ref{eq:rho_peak_val}) and thus totally unrelated to the data used during the training.}\label{fig:peaks_sig}
\end{figure}
In order to validate our method we repeated the test just described 2000 times. We have generated random spectral densities, not used in the trainings, by starting again from Eq.~(\ref{eq:rho_cheby}) and by selecting random values for $E_0$ and $N_\mathrm{b}$ respectively in the intervals  $[0.3,1.3]$ GeV and $[8,1024]$. 

The values of $p_\sigma(E)$ for all the energies and for all the samples is collected in one set, $p_\sigma$ for short, whose normalized distribution is shown in the left panels of FIG.~\ref{fig:cheby_sig} (top  panel for  $\sigma=0.44$~GeV, bottom panel for $\sigma=0.63$~GeV). The distributions are nicely bell-shaped around 0 and the right panels in FIG.~\ref{fig:cheby_sig} show the comparison between the normalized cumulative distribution functions of $p_\sigma$ with the normal distribution obtained from the mean and variance of $p_\sigma$.  The p-value calculated from the Kolmogorov-Smirnov test is much less than 0.05 in both the cases and therefore the distribution of $p_\sigma$ cannot be considered as a normal one. On the other hand, given the procedure that we use to quote the total error, we observe deviations $\hat{\rho}_\sigma^\mathrm{pred}(E)-\hat{\rho}_\sigma^\mathrm{true}(E)$ smaller than 2 standard deviations in 95\% of the cases and smaller than 3 standard deviations in 99\% of the cases. The fact that deviations smaller than 1 standard deviation occur in $\sim$80\% of the cases for $\sigma=0.44$ GeV and $\sim$85\% of the cases for $\sigma=0.63$ GeV (to be compared with the expected 68\% in the case of a  normal distribution) is an indication of the fact that our estimate of the total error is indeed conservative.  Moreover, from FIG.~\ref{fig:cheby_sig} it is also evident that our ensembles of neural networks are able to generalize very efficiently outside the training set.

In order to perform another stringent validation test of the proposed method, we also considered unsmeared spectral densities that cannot be written on the  Chebyshev basis. We repeated the analysis described in the previous paragraph for a set of spectral densities generated according to
\begin{flalign}\label{eq:rho_peak_val}
	\rho(E)=\sum_{n=1}^{N_\mathrm{peaks}} c_n \delta(E-E_n),
\end{flalign}
where $0<E_0\le E_1 \le E_2 \le \cdots \le E_\mathrm{peaks}$. By plugging Eq.~(\ref{eq:rho_peak_val}) into Eq.~(\ref{eq:inputC}) and Eq.~(\ref{eq:outputRho}) we have calculated the correlator and smeared spectral density associated to each unsmeared $\rho(E)$. We have generated 2000 unsmeared spectral densities with $N_\mathrm{peaks}=5000$. The position $E_n$ of each peak has been set by drawing independent random numbers uniformly distributed in the interval $[E_0,15\,\mathrm{GeV}]$ while $E_0$ has been randomly chosen in the interval $[0.3,1.3]$ GeV. The coefficients $c_n$ have also been generated randomly in the interval $[-0.01,0.01]$. The 2000 trains of isolated Dirac-delta peaks are representative of unsmeared spectral densities that might arise in the study of finite volume lattice correlators. These are rather wild and irregular objects that our neural networks have not seen during the trainings. FIG~\ref{fig:peaks_sig} shows the plots equivalent to those of FIG~\ref{fig:cheby_sig} for this new validation set. As it can be seen, the distributions of $p_\sigma(E)$ are basically unchanged, an additional reassuring evidence of the ability of our ensembles of neural networks to generalize very efficiently outside the training set and, more importantly, on the robustness of the procedure that we use to estimate the errors.

\section{\label{sec:mock_data}Results for mock data inspired by physical models}

In the light of the results of the previous section, providing a solid quantitative evidence of the robustness and reliability of the proposed method, we investigate in this section the performances of our trained ensembles of machines in the case of mock spectral densities coming from physical models. A few more validation tests, unrelated to physical models, are discussed in Appendix~\ref{app:examples}. 

For each test discussed in the following subsections we define a model spectral density $\rho(E)$ and then use Eq.~(\ref{eq:inputC}) and Eq.~(\ref{eq:outputRho}) to calculate the associated exact correlator and true smeared spectral density. We then generate $N_\mathrm{c}=800$ bootstrap samples $C_c(t)$, by sampling the distribution of Eq.~(\ref{eq:Multivariate}), and quote our final results by using the procedure described in details in the previous section.

\subsection{\label{subsec:isolated_multi}A resonance and a multi-particle plateaux}
\begin{table}[t!]
	\begin{ruledtabular}
		\begin{tabular}{cccccc}
			Model & $C$ & $M$ [GeV]  & $\Gamma$ [GeV] & $\delta_2$ [GeV]   \\  \hline
			$\rho^\mathrm{GM}_1(E)$ & 10 & 1 & 0.1 & 1.5\\ 
			$\rho^\mathrm{GM}_2(E)$ & 5  & 1 & 0.1 & 1.5\\ 
			$\rho^\mathrm{GM}_3(E)$ & 5  & 0.8 & 0.05 & 1.2\\ 
		\end{tabular}
	\end{ruledtabular}
	\caption{\small	Additional parameters used to generate the mock unsmeared spectral densities based on a Gaussian mixture model, Eq.~\ref{eq:GM}. The corresponding functions are plotted in FIG.~\ref{fig:res_gaussian_mixture} (dashed grey). }\label{tab:GMparam} 
\end{table}
\begin{figure}[t!]
	\includegraphics[width=\columnwidth]{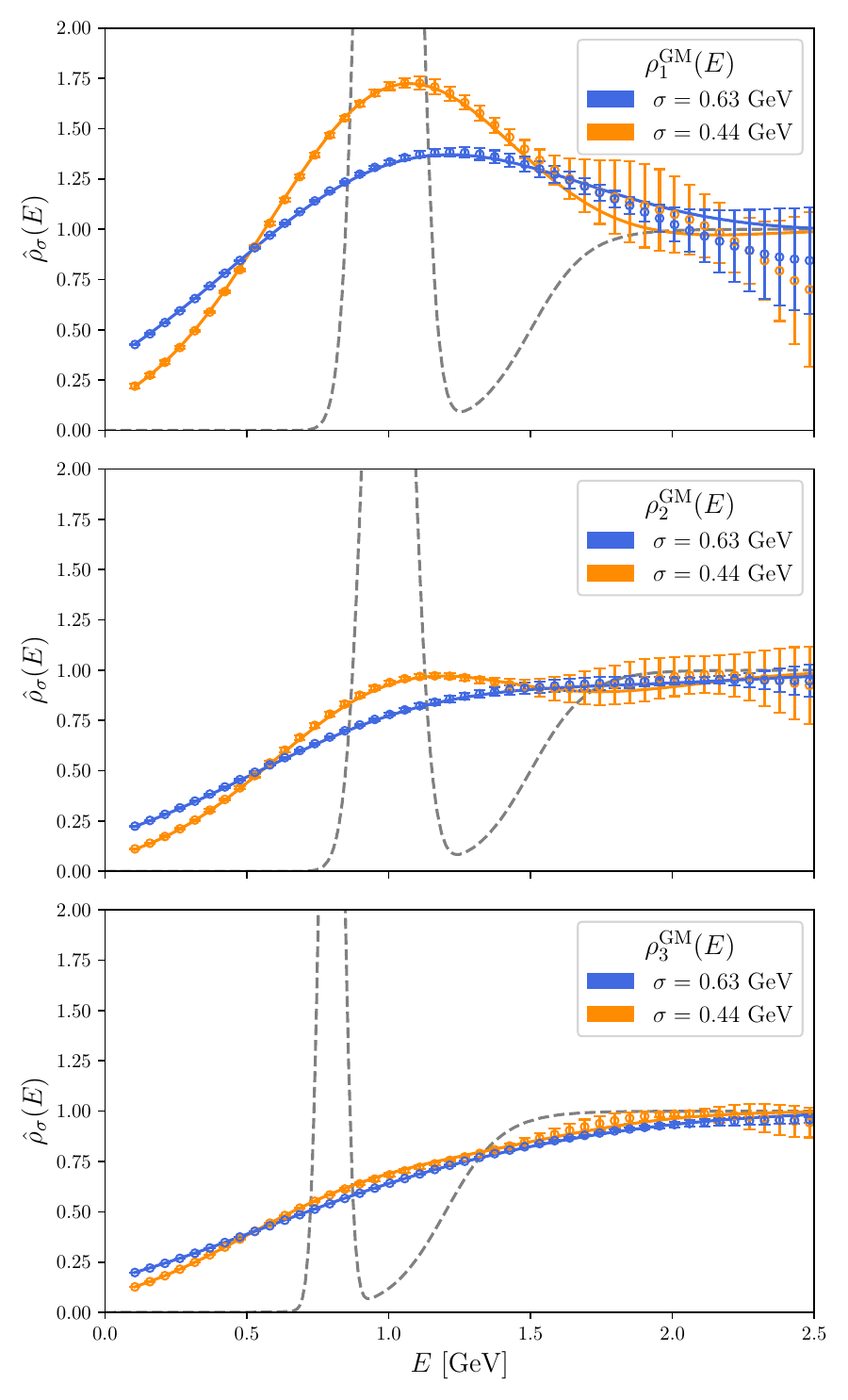}
	\caption{\small Results for three different unsmeared spectral densities $\rho(E)$ (dashed grey lines) generated from the Gaussian mixture model of Eq.~(\ref{eq:GM}). The solid lines correspond to $\hat{\rho}_\sigma^\mathrm{true}(E)$. The points, with error bars, refer to the predicted spectral densities $\hat{\rho}_\sigma^\mathrm{pred}(E)$.}
	\label{fig:res_gaussian_mixture}
\end{figure}
\begin{figure}[t!]
	\includegraphics[width=\columnwidth]{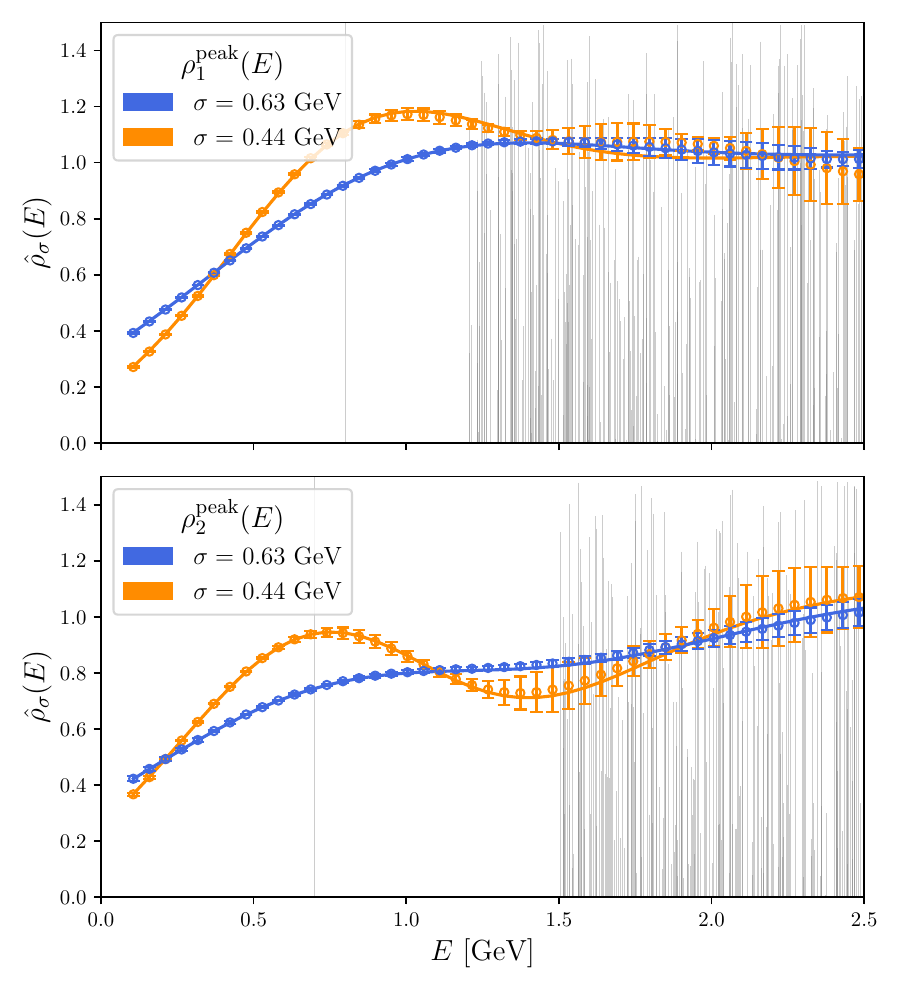}
	\caption{\small Reconstructed smeared spectral densities (points with errors) compared to the true ones (solid lines) for two unsmeared spectral densities  simulating a finite volume distribution, see Eq.~(\ref{eq:1peak}). The unsmeared $\rho(E)$ is represented by vertical lines located in correspondence of $E=E_n$ with heights proportional to $c_n$.}
	\label{fig:res_peaks}
\end{figure}
The first class of physically motivated models that we investigate is that of unsmeared spectral densities exhibiting the structures corresponding to an isolated resonance and a multi-particle plateau. To build mock spectral densities belonging to this class we used the same Gaussian mixture model used in Ref.~\cite{Chen:2021giw} and given by
\begin{flalign}\label{eq:GM}	
	\rho^\mathrm{GM}(E)
	&=\hat{\theta}(E,\delta_1,\zeta_1) \left[C e^{-\left(\frac{E-M}{\Gamma}\right)^2} \left(1-\hat{\theta}(E,\delta_2,\zeta_2)\right)  \right]
	\nonumber 	\\
	\nonumber 	\\
	&
	+ C_0 \hat{\theta}(E,\delta_1,\zeta_1)\hat{\theta}(E,\delta_2,\zeta_2),
\end{flalign}  
where
\begin{flalign}
	\hat{\theta}(E,\delta_i,\zeta_i) = \frac{1}{1+\exp \left(-\frac{E-\delta_i}{\zeta_i}\right)}.
\end{flalign}
The sigmoid function 
$\hat{\theta}(E,\delta_1,\zeta_1)$ with $\delta_1=0.1$ GeV and $\zeta_1=0.01$ GeV dumps $\rho^\mathrm{GM}(E)$ at low energies while the other sigmoid $\hat{\theta}(E,\delta_2,\zeta_2)$,  with $\zeta_2=0.1$ GeV, connects the resonance to the continuum part whose threshold is $\delta_2$. 
The parameter $C_0=1$ regulates the height of the continuum plateaux which also coincides with the asymptotic behaviour of the spectral density, i.e. $\rho^\mathrm{GM}(E)\mapsto C_0$ for $E\mapsto \infty$. We have generated three different spectral densities with this model that, in the following, we call $\rho_1^\mathrm{GM}(E)$, $\rho_2^\mathrm{GM}(E)$ and $\rho_3^\mathrm{GM}(E)$ and whose parameters are given in TABLE~\ref{tab:GMparam}.

The predicted smeared spectral densities for smearing widths $\sigma=0.44$ GeV and $\sigma=0.63$ GeV are compared to the true ones in FIG.~\ref{fig:res_gaussian_mixture}. In all cases the predicted result agrees with the true one,  within the quoted errors, for all the explored values of $E$. The quality of the reconstruction of the smeared spectral densities is excellent for $E<1.5$ GeV while, at higher energies, the quoted error is sufficiently large to account for the deviation of $\hat{\rho}_\sigma^\mathrm{pred}(E)$ from $\hat{\rho}_\sigma^\mathrm{true}(E)$. This is particularly evident in the case of $\rho_1^\mathrm{GM}(E)$ shown in the top panel.

The same class of physical models can also be investigated by starting from unsmeared spectral densities that might arise in finite volume calculations by considering
\begin{flalign}\label{eq:1peak}
\rho^\mathrm{peak}(E) = C_\mathrm{peak}\,\delta(E-E_\mathrm{peak}) + \sum_{n=1}^{N_\mathrm{peaks}} c_n \delta(E-E_n).
\end{flalign}
The parameter $E_\mathrm{peak}$ parametrizes the position of an isolated peak while the multi-particle part is introduced with the other peaks located at $E_\mathrm{peak}<E_1\le \cdots \le E_\mathrm{N_\mathrm{peaks}}$. We have generated two unsmeared spectral densities, $\rho_1^\mathrm{peak}(E)$ and $\rho_2^\mathrm{peak}(E)$ by using this model. In both cases we set $N_\mathrm{peaks}=10000$, selected random values for the $E_n$'s up to $50$ GeV and random values for the $c_n$'s in the range $[0,0.01]$. In the case of $\rho_1^\mathrm{peak}(E)$ we set $E_\mathrm{peak}=0.8$ GeV, $C_\mathrm{peak}=1$ and $E_1=1.2$ GeV. In the case of $\rho_2^\mathrm{peak}(E)$ we set instead $E_\mathrm{peak}=0.7$ GeV, $C_\mathrm{peak}=1$ and $E_1=1.5$ GeV.

The predicted smeared spectral densities are compared with the true ones in FIG.~\ref{fig:res_peaks}. In both cases $\hat{\rho}_\sigma^\mathrm{pred}(E)$ is in excellent agreement with $\hat{\rho}_\sigma^\mathrm{true}(E)$.

It is worth emphasizing once again that the model in Eq.~(\ref{eq:1peak}) cannot be represented by using the Chebyshev basis of Eq.~(\ref{eq:rho_cheby}) and, therefore, it is totally unrelated to the smooth unsmeared spectral densities that we used to populate the training sets.

\subsection{$O(3)$ non-linear $\sigma$-model}
\begin{figure}[t!]
	\includegraphics[width=\columnwidth]{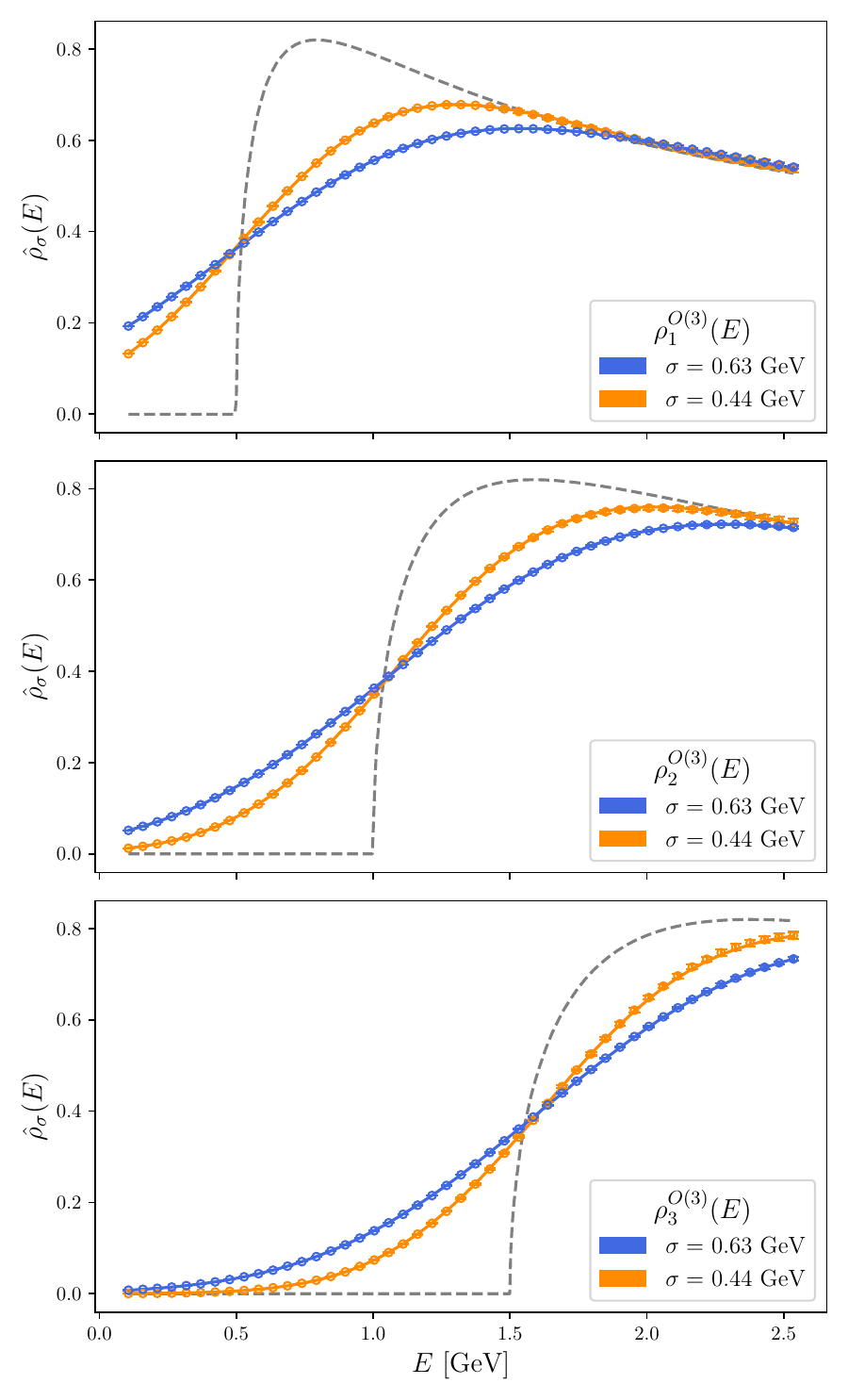}
	\caption{\small Predicted smeared spectral densities for the two-particles contribution to the vector-vector correlator in the $1+1$ dimensional $O(3)$ non-linear $\sigma$ model for different values of the two-particle threshold: 0.5 GeV (top panel), 1 GeV (central panel) and 1.5 GeV (third panel). The solid lines and the dashed curve correspond respectively to $\hat{\rho}_\sigma^\mathrm{pred}(E)$ and $\rho^{O(3)}(E)$.}
	\label{fig:res_O3}
\end{figure}
In this subsection we consider a model for the unsmeared spectral density that has already been investigated by using the HLT method in Ref.~\cite{Bulava:2021fre}. More precisely, we consider the two-particles contribution to the vector-vector spectral density in the the $1+1$ dimensional $O(3)$ non-linear $\sigma$-model (see Ref.~\cite{Bulava:2021fre} for more details) given by
\begin{flalign}
&
\rho^{O(3)}(E) 
\nonumber \\
&
= \theta(E-E_\mathrm{th})\frac{3\pi^3}{4\theta^2} \frac{\theta^2+\pi^2}{\theta^2+4\pi^2} \tanh^3 \frac{\theta}{2} \bigg\vert_{\theta=2 \cosh^{-1}\frac{E}{E_\mathrm{th}}},
\end{flalign}
where $E_\mathrm{th}$ is the two-particle threshold. We considered three mock unsmeared spectral densities, that we call $\rho^{O(3)}_1(E)$, $\rho^{O(3)}_2(E)$ and $\rho^{O(3)}_3(E)$, differing for the position of the multi-particle threshold, which has been set respectively to $E_\mathrm{th}=0.5$ GeV, 1 GeV and 1.5 GeV. The reconstructed smeared spectral densities for $\sigma=0.44$ GeV and $\sigma=0.63$ GeV are compared with the exact ones in FIG.~\ref{fig:res_O3}.

The predicted smeared spectral densities are in remarkably good agreement with the true ones in the full energy range and in all cases.  This result can be read as an indication of the fact that the smoothness of the underlying unsmeared spectral density plays a crucial r\^ole in the precision that one can get on $\hat{\rho}_\sigma^\mathrm{pred}(E)$, also if the problem is approached by using a neural network approach. Indeed, this fact had already been observed and exploited in Ref.~\cite{Bulava:2021fre}, where the authors managed to perform the $\sigma\mapsto 0$ limit of $\hat{\rho}_\sigma^{O(3)}(E)$ with controlled systematic errors by using the HLT method.

\subsection{\label{sec:Rratio}The $R$-ratio with mock data}
\begin{figure}[t!]
	\includegraphics[width=\columnwidth]{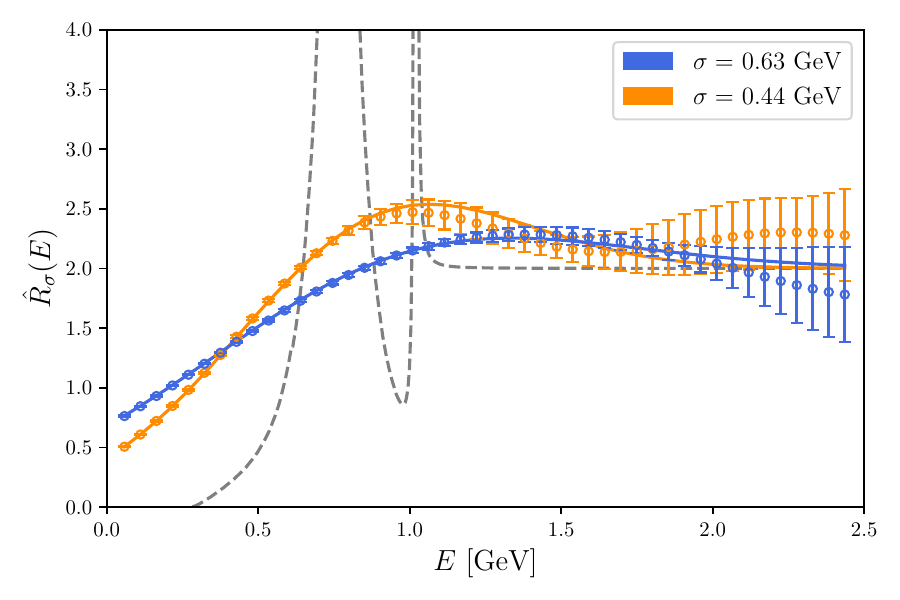}
	\caption{\small Results for the $R$-ratio using the parametrization of Ref.~\cite{Bernecker:2011gh}. The gray dashed line is $R(E)$ while the solid lines correspond to $\hat{R}_\sigma^\mathrm{true}(E)$.}
	\label{fig:res_R}
\end{figure}
The last case that we consider is that of a model spectral density coming from a parametrization of the experimental data of the  $R$-ratio. The $R$-ratio, denoted by $R(E)$, is defined as the ratio between the inclusive cross section of $e^+e^-\to \mathrm{hadrons}$ and $e^+e^-\to \mu^+\mu^-$ and plays a crucial r\^ole in particle physics phenomenology (see e.g. Refs.~\cite{Aoyama:2020ynm,ExtendedTwistedMassCollaborationETMC:2022sta}).

In the next section we will present results for a contribution to the smeared $R$-ratio $R_\sigma(E)$ that we obtained by feeding into our ensembles of trained neural networks a lattice QCD correlator that has been already used in Ref.~\cite{ExtendedTwistedMassCollaborationETMC:2022sta} to calculate the same quantity with the HLT method.

Before doing that, however, we wanted also to perform a test with mock data generated by starting from the parametrization of $R(E)$ given in Ref.~\cite{Bernecker:2011gh}. This parametrization, that we use as model unsmeared spectral density by setting $\rho(E)\equiv R(E)$, is meant to reproduce the experimental measurements of $R(E)$ for energies $E<1.1$~GeV, i.e.\ in the low-energy region where there are two dominant structures associated with the mixed $\rho$ and $\omega$ resonances, a rather broad peak at $E\simeq 0.7$~GeV, and the narrow resonance $\phi(1020)$. Given this rich structure, shown as the dashed grey curve in FIG.~\ref{fig:res_R}, this is a much more challenging validation test w.r.t.\ the one of the $O(3)$ model discussed in the previous subsection.

In FIG.~\ref{fig:res_R}, $\hat{R}_\sigma^\mathrm{pred}(E)$ is compared with $\hat{R}_\sigma^\mathrm{true}(E)$ for both $\sigma=0.44$~GeV (orange points and solid curve) and  $\sigma=0.63$~GeV (blue points and solid curve). In both cases the difference  $\hat{R}_\sigma^\mathrm{pred}(E)-\hat{R}_\sigma^\mathrm{true}(E)$ doesn't exceed the quoted error for all the considered values of $E$.

\section{\label{sec:lattice_data}The $R$-ratio with lattice QCD data}
\begin{table}[t]
	\begin{ruledtabular}
		\begin{tabular}{lcccc}
			\textrm{ID}&
			$L^3\times T$&
			$a$ \textrm{fm}&
			$aL$ \textrm{fm}&
			$m_\pi$ \textrm{GeV}\\
			\colrule
			\textrm{B64} & $64^3\times 128$ & 0.07957(13) & 5.09 & 0.1352(2) 
		\end{tabular}
	\end{ruledtabular}
	\caption{\label{tab:ensembles}%
		ETMC gauge ensemble used in this work. See Refs.~\cite{Alexandrou:2022amy,ExtendedTwistedMassCollaborationETMC:2022sta} for more details.
	}
\end{table}
\begin{figure}[t!]
	\includegraphics[width=\columnwidth]{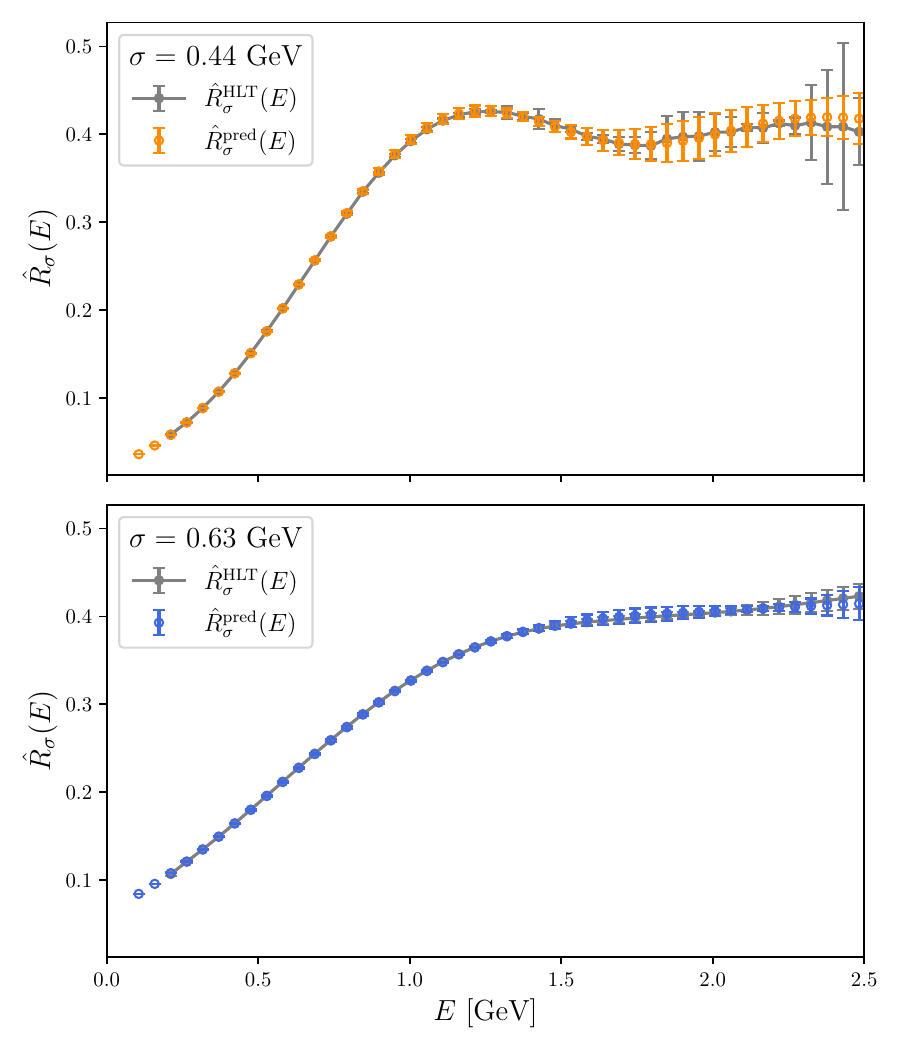}
	\caption{\small The results obtained with our ensembles of neural networks in the case of a real lattice QCD correlator are compared with those obtained with the HLT method (grey, Ref.~\cite{ExtendedTwistedMassCollaborationETMC:2022sta,Hansen:2019idp}). The top-panel shows the two determinations of the strange-strange connected contribution to the smeared $R$-ratio for $\sigma=0.44$~GeV while the case $\sigma=0.63$~GeV is shown in the bottom-panel.}
	\label{fig:res_lattice_ss1}
\end{figure}
\begin{figure}[h!]
	\includegraphics[width=\columnwidth]{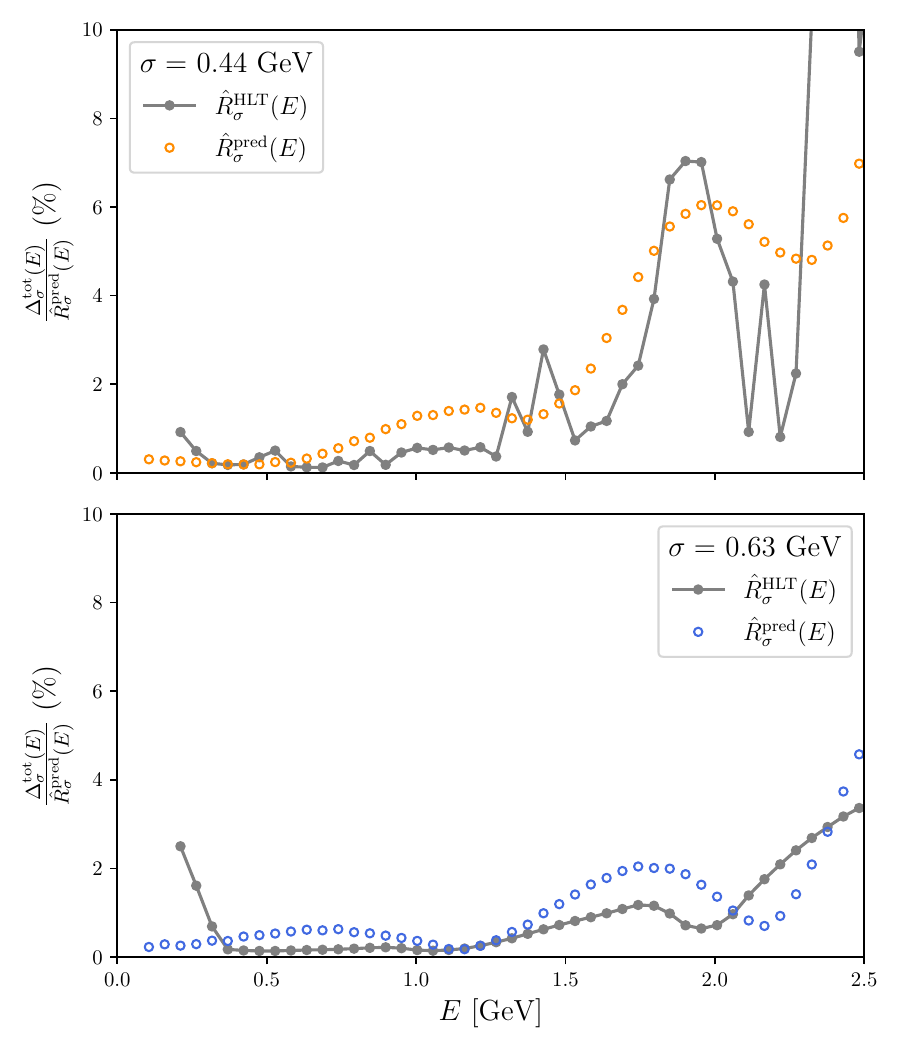}
	\caption{\small Relative error on the smeared spectral density obtained from our ensembles of neural networks (colored points) and the HLT method (grey points) for $\sigma=0.44$ GeV (top panel) and $\sigma=0.63$ GeV (bottom panel).}
	\label{fig:relative_error_ss1}
\end{figure}
In this section we use our trained ensembles of neural networks to extract the smeared spectral density from a real lattice QCD correlator.  

We have considered a lattice correlator, measured by the ETMC on the ensemble described in TABLE~\ref{tab:ensembles}, that has already been used in Ref.~\cite{ExtendedTwistedMassCollaborationETMC:2022sta} to extract the so-called strange-strange connected contribution to the smeared $R$-ratio by using the HLT method of Ref.~\cite{Hansen:2019idp}. The choice is motivated by the fact that in this case the exact answer for the smeared spectral density is not known and we are going to compare our results $\hat{R}_\sigma^\mathrm{pred}(E)$ with the ones, that we call $\hat{R}_\sigma^\mathrm{HLT}(E)$, obtained in Ref.~\cite{ExtendedTwistedMassCollaborationETMC:2022sta}.

In QCD the $R$-ratio can be extracted from the lattice correlator 
\begin{flalign}\label{eq:Clattice}
	C_\mathrm{latt}(t)
	&=-\frac{1}{3}\sum_{i=1}^3\int \mathrm{d}^3x \,\mathrm{T}\bra{0}J_i(x)J_i(0)\ket{0}
	\nonumber \\
	\nonumber \\
	&=\int_{0}^\infty \mathrm{d}\omega \, \frac{\omega^2}{12\pi^2}e^{-t\omega} R(\omega)\;.
\end{flalign}
where $J_\mu(x)$ is the hadronic electromagnetic current. The previous formula (in which we have neglected the term proportional to $e^{-(T-t)\omega}$ that vanishes in the $T\mapsto \infty$ limit) explains the choice that we made in Eq.~(\ref{eq:inputC}) to represent all the correlators that we used as inputs to our neural networks. In Ref.~\cite{ExtendedTwistedMassCollaborationETMC:2022sta} all the connected and disconnected contributions to $C_\mathrm{latt}(t)$, coming from the fact that $J_\mu=\sum_{f}q_f \bar \psi_f \gamma_\mu \psi_f$ with $f=\{u,d,s,c,b,t\}$, $q_{u,c,t}=2/3$ and $q_{d,s,b}=-1/3$, have been taken into account. Here we consider only the connected strange-strange contribution, i.e.\ we set $J_\mu = -\frac{1}{3} \bar{s} \gamma_\mu s$ and neglect the contribution associated with the fermionic-disconnected Wick contraction. From $C_\mathrm{latt}(t)$ we have calculated the covariance matrix $\hat{\Sigma}_\mathrm{latt}$ that we used to inject noise in all the training sets that we have built and used in this work (see Section~\ref{sec:noise}). 

Although, as already stressed, in this case we don't know the exact smeared spectral density, we do expect from phenomenology to see a sizeable contribution to the unsmeared spectral density coming from the $\phi(1020)$ resonance. Therefore, the shape of the smeared spectral density is expected to be similar to those shown in FIG.~\ref{fig:res_peaks} for which the results obtained from our ensembles of neural networks turned out to be reliable. The comparison of $\hat{R}_\sigma^\mathrm{pred}(E)$ and $\hat{R}_\sigma^\mathrm{HLT}(E)$ is shown in FIG.~\ref{fig:res_lattice_ss1} and, as it can be seen, the agreement between the two determinations is remarkably good.

Some remarks are in order here. The HLT method and the new method that we propose here are radically different and the fact that the results obtained with the two procedures are in such a good agreement, especially for $E<1.5$ GeV where both methods provide very small errors, is at the same time reassuring and encouraging in view of future applications of spectral reconstruction techniques. 
 
Concerning the errors, there is no significant evidence that one method has better performances than the other. This can be better  appreciated in FIG.~\ref{fig:relative_error_ss1} where the relative error is shown in both cases as a function of the energy. 

We want to stress that in the new proposed method, as in the HLT case, no prior information on the expected spectral density has to be provided and, therefore, the results for $\hat{R}_\sigma^\mathrm{pred}(E)$ shown in FIG.~\ref{fig:res_lattice_ss1} have to be considered as first-principles model-independent determinations of the smeared strange-strange contribution to the $R$-ratio that we managed to obtain by using supervised deep-learning techniques. 

We leave to future work on the subject the task of analysing all the contributions to $R(E)$, as done in Ref.~\cite{ExtendedTwistedMassCollaborationETMC:2022sta}, in order to obtain a machine-learning determination of $\hat{R}_\sigma(E)$ to be compared with experiments.
 
\section{Conclusions} \label{sec:conclusions}
In this work we have proposed a new method to extract smeared hadronic spectral densities from lattice correlators. The method has been built by using supervised deep-learning techniques and is characterized by the distinctive features to implement a model-independent training strategy and to provide a reliable estimate of both the statistical and systematic uncertainties.

We managed to implement a model-independent training strategy by introducing a basis in the functional space from which we extract the spectral densities that we use to populate our training sets. In order to obtain a reliable estimate of the systematic errors, we introduced the ensembles of machines. 

We have shown that, by studying the distribution of the answers of the different machines belonging to an ensemble, at fixed and finite number of neurons, dimension and complexity of the training set, it is possible to quantify reliably the systematic errors associated with the proposed method. To do that, we presented a large number of stringent validation tests, performed with mock data, providing quantitative evidence of the reliability of the procedure that we use to estimate the total error on our final results (see FIG.~\ref{fig:cheby_sig} and FIG.~\ref{fig:peaks_sig}).

In addition to mock data, we have applied the new proposed method also in the case of a lattice QCD correlator obtained from a simulation performed by the ETM Collaboration. We have extracted the strange-strange connected contribution to the smeared $R$-ratio and compared the predictions obtained by using our ensembles of trained machines with the ones previously obtained by using the HLT method~\cite{ExtendedTwistedMassCollaborationETMC:2022sta,Hansen:2019idp}. We found a remarkably good agreement between the results obtained by using the two totally unrelated methods that provide total errors of the same order of magnitude.

The proposed method requires the training of many machines with different architectures and dimensions of the training sets. Admittedly, although this is a task that can be completed in a few days on a modern desktop computer, the procedure might end up to be computationally demanding. On the one hand, we don't exclude the possibility that the proposed method can be simplified in order to speed up the required computations. At the same time, on the basis of our experience, we are firmly convinced that a careful study of the different sources of systematic uncertainties is mandatory when dealing with machine-learning techniques and when the aim is to compare theoretical predictions with experiments. In fact, the computational cost of the proposed method is the price that we had to pay for the reliability of the results.

As a final remark we want to stress that in this paper we have taught a lesson to a broad audience of learning-machines. The subject of the lesson, the extraction of smeared spectral densities from lattice correlators, is just a particular topic. The idea of teaching \emph{systematically} to a broad audience of machines is much more general and can be used to estimate reliably the systematic errors in many other applications.

\acknowledgements{}

We warmly thank our colleagues of the ETMC for very useful discussions and for providing us the data of the lattice QCD correlator used in this study. We also thank L.~Del Debbio, R.~Frezzotti, A.~Lupo, M.~Panero and M.~Sbragaglia for enlightening discussions and B.~Lucini for his extremely valuable comments on a preliminary version of the manuscript. MB acknowledges support from the European Research Council (ERC) under the European Union’s Horizon 2020 research and innovation programme (Grant Agreement No. 882340).

\appendix

\section{\label{app:examples} Results in exceptional cases}

\begin{figure}[]
	\includegraphics[width=\columnwidth]{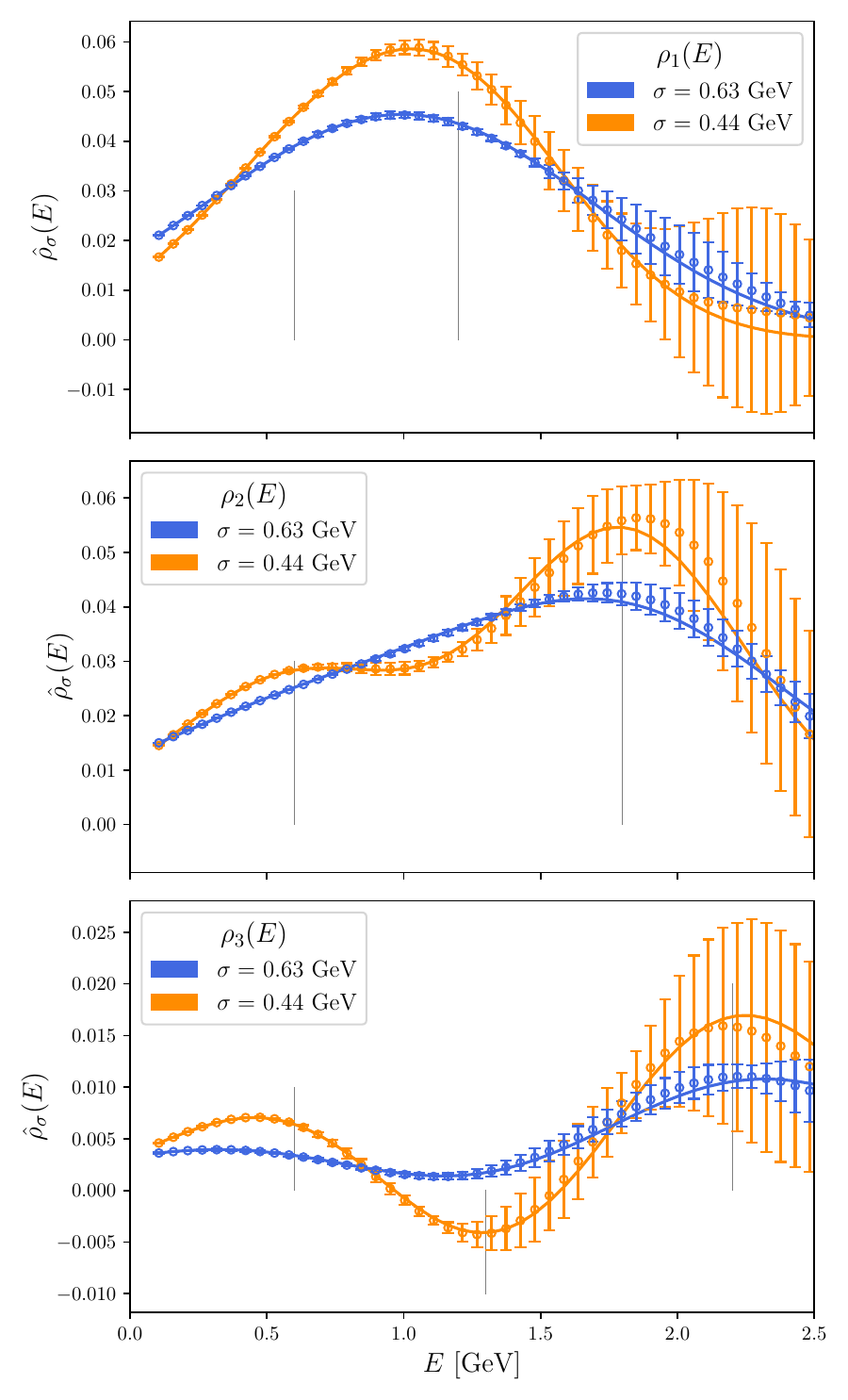}
	\caption{\small Predicted smeared spectral densities for the three models of unsmeared spectral densities given in Eq.~(\ref{eq:threepeacks}). The solid lines correspond to $\hat{\rho}_\sigma^\mathrm{true}(E)$. The vertical grey lines are in correspondence of the isolated Dirac-delta peaks and the height is proportional to the associated weights.}
	\label{fig:res_isolated_peaks}
\end{figure}
In this appendix we illustrate some additional validation tests that we performed by using mock data designed in order to challenge our ensembles of trained machines in extreme situations, i.e.\ in the case of unsmeared spectral densities that are very different from the ones used in the trainings. To this end, we considered the following unsmeared spectral densities 
\begin{flalign}
&
\rho_1(E)= 0.03\cdot \delta(E-0.6) +  0.05\cdot\delta(E-1.2)\;,
\nonumber \\
\nonumber \\
&
\rho_2(E)= 0.03\cdot \delta(E-0.6) +  0.06\cdot\delta(E-1.8)\;,
\nonumber \\
\nonumber \\
&
\rho_3(E)=  0.01\cdot \delta(E-0.6) -0.01\cdot\delta(E-1.3)
\nonumber \\
&\qquad\qquad+0.02\cdot \delta(E-2.2)\;,
\label{eq:threepeacks}
\end{flalign}
where the arguments of the Dirac-delta functions are in GeV units. The above spectral densities correspond to either two or three very well separated Dirac-delta peaks and one of the peaks of $\rho_3(E)$ has a negative coefficient. The final predicted results, compared with the exact ones, are shown in FIG.~\ref{fig:res_isolated_peaks} for $\sigma=0.44$ GeV and $\sigma=0.63$ GeV. The agreement with $\hat{\rho}_\sigma^\mathrm{true}(E)$ is excellent in all cases. As it can be seen, a difference w.r.t.\ the other models considered in Sections~\ref{sec:quote_results} and~\ref{sec:mock_data} is the size of the errors that are much larger in these cases for $\sigma=0.44$ GeV. This 
had to be expected and is a desired feature. Indeed, the inverse problem becomes harder when one tries to reconstruct sharp peaks with high resolution. The presence of statistical noise in the input data prevents any method to provide a very precise result, especially at high energies. Our neural network strategy, in these exceptional situations, provides a large error and this is a further reassuring evidence of the robustness of the proposed procedure.

\section{\label{sec:input_time_slices}Dependence on the number of input time slices}

\begin{figure}[t!]
	\includegraphics[width=\columnwidth]{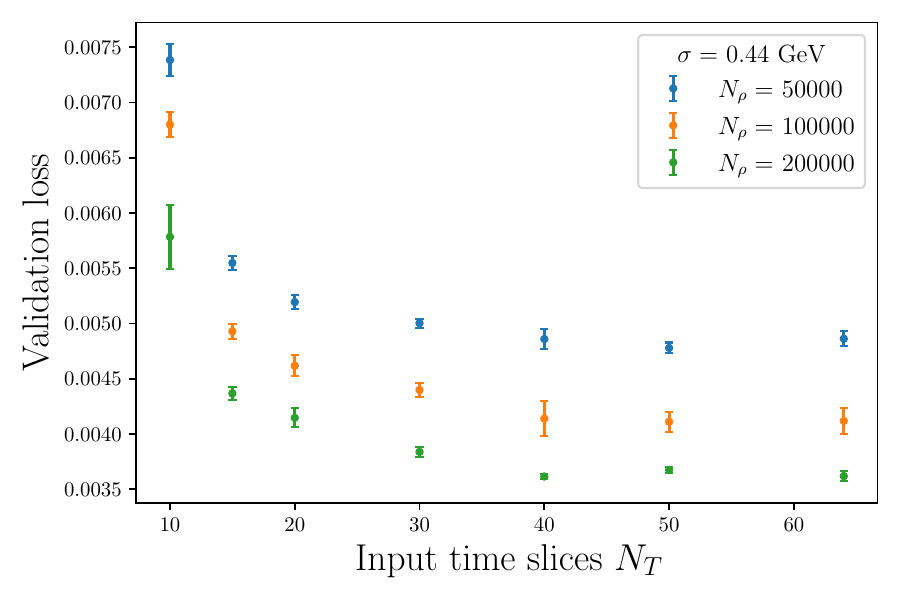}
	\caption{\small Validation losses at the end of the training for different number of time slices in the input data and for different training set size $N_\rho$. This test is performed with $N_\mathrm{b}=512$ and $\sigma=0.44$ GeV. Each point with the associated error bar comes from the average of $N_\mathrm{r}=20$ trainings (see Section~\ref{sec:training_strategy}). }
	\label{fig:input_time_slices}
\end{figure}
In Section~\ref{sec:setup} we emphasized that the noise-to-signal ratio of generic lattice correlators grows exponentially with the Euclidean time (see FIG.~\ref{fig:SN_obs_B64}) and that for this reason it might be convenient to reduce the number $N_T$ of components of the input vector $\vec C$.

The approaches based on the Backus-Gilbert regularization method have a built–in mechanism to suppress noisy input data and by using these methods one can safely feed the whole information contained in the correlator into the algorithm. In general, this cannot be expected to be true when supervised deep-learning approaches are employed. Even though neural networks have been proven to be robust in handling noisy data, too much noise can have a bad impact on the performances since a big fraction of the effort in the minimization algorithm is put in the suppression of the noise and in learning the distinction between outliers and effective information rather than in learning general features of the problem. 

In the following we investigate the dependence of the training performances, obtained by injecting the noise of the lattice correlator in the training sets as explained in subsection~\ref{sec:noise}, upon $N_T$. In FIG.~\ref{fig:input_time_slices} we show the validation loss as a function of $N_T$. As it can be seen, the performances of the networks improve as $N_T$ increases but the validation loss reaches a saturation point around $N^\mathrm{sat}_T=40$. On the one hand, $N^\mathrm{sat}_T$ can be considered as the maximum number of time slices of the correlator from which meaningful physical information can be extracted. On the other hand, despite the corruption of the input data for $N_T>N^\mathrm{sat}_T$, including more time slices does not compromise the quality of the trainings and we could safely set $N_T=T/a=64$. 

%

\section{\label{sec:netVSlatt}Analysis of the statistical uncertainties}
\begin{figure}[t!]
	\includegraphics[width=\columnwidth]{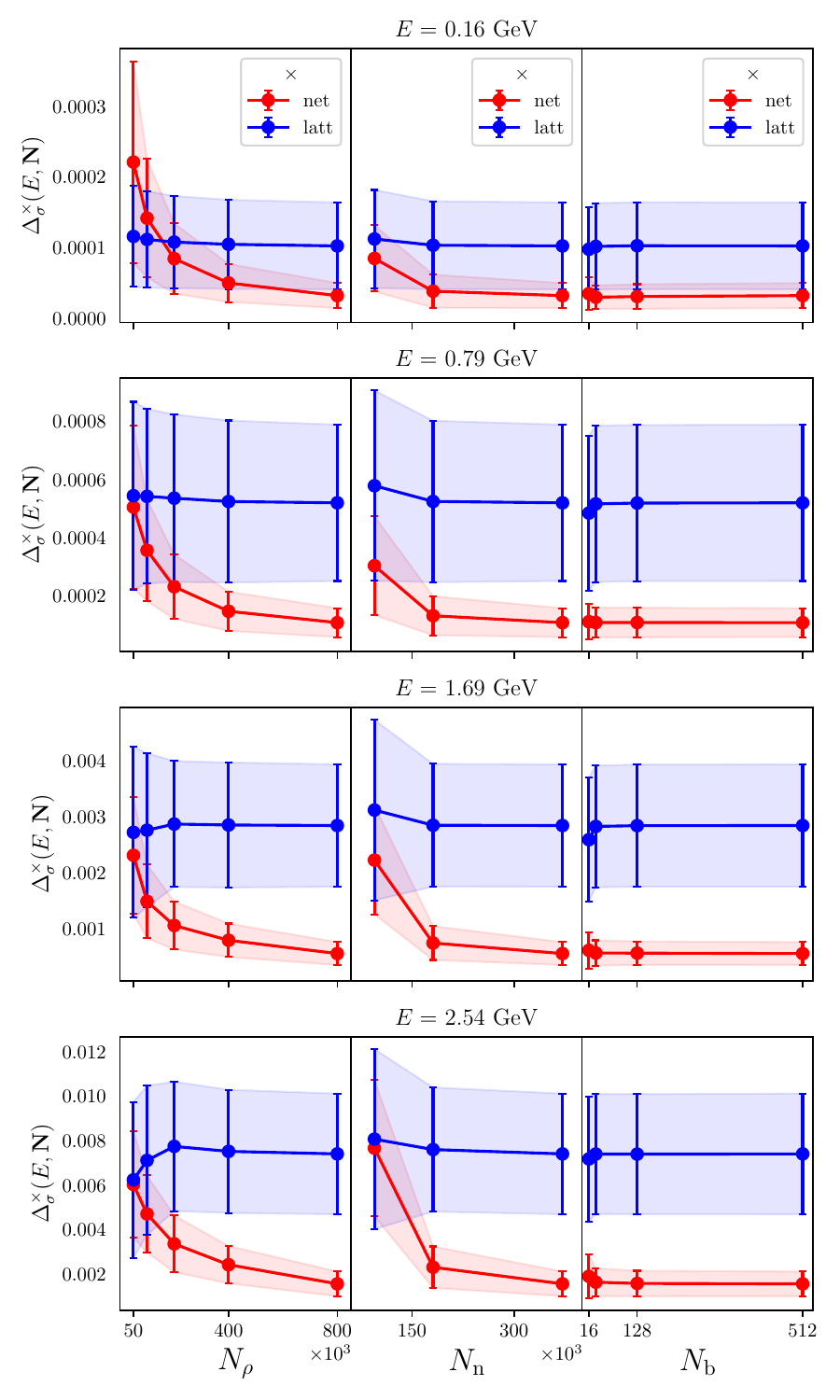}\caption{\small Behaviour of $\Delta_\sigma^\mathrm{latt}(E,\vec{N})$ (blue) and $\Delta_\sigma^\mathrm{net}(E,\vec{N})$ (red)  with respect to $\vec{N}=\{N_\rho,N_\mathrm{n},N_\mathrm{b}\}$ at four different values of the energy. The data refer to $\sigma=0.44$ GeV and have been obtained by analysing the 2000  samples built on the Chebyshev basis already used in Section~\ref{sec:velidation}. The central values and the associated error bars come from the average over the 2000 samples.}\label{fig:netVSlatt}
\end{figure}
In illustrating our method we have stressed that there are two sources of statistical errors that contribute to $\Delta_\sigma^\mathrm{stat}(E,\vec{N})$ defined in  Eq.~(\ref{eq:stat}): $\Delta_\sigma^\mathrm{latt}(E,\vec{N})$, coming from the Monte Carlo simulation, and $\Delta_\sigma^\mathrm{net}(E,\vec{N})$, coming from our ensemble of trained machines. In this appendix we study the dependence of $\Delta_\sigma^\mathrm{latt}$ and $\Delta_\sigma^\mathrm{net}$ upon the energy $E$ and upon $\vec{N}$. The investigation is carried out by collecting $\Delta_\sigma^\mathrm{latt}$  and $\Delta_\sigma^\mathrm{net}$ for the 2000 mock samples built on the Chebyshev basis and used in Section~\ref{sec:velidation} to validate our procedure. For each combination of $\vec{N}$ the 2000 results are averaged and the standard deviation is calculated. The results are shown in FIG.~\ref{fig:netVSlatt} at $\sigma=$~0.44 GeV and at four different energies spanning the output energy range. 
We observe that both $\Delta_\sigma^\mathrm{latt}$ and $\Delta_\sigma^\mathrm{net}$  increase in magnitude as the energy increases. This phenomenology is a well-known feature of the inverse Laplace transform problem in which  the contributions to the correlation function coming from high energies are strongly suppressed by the exponential basis (see Eq.~(\ref{eq:CfiniteL}).  Therefore, even small statistical fluctuations in the correlator turn into enhanced errors in the spectral density at high energy. 
At fixed energy the Monte Carlo error $\Delta_\sigma^\mathrm{latt}$ does not show a significant dependence upon $\vec{N}$. This is a reassuring evidence that $\Delta_\sigma^\mathrm{latt}$ and $\Delta_\sigma^\mathrm{net}$ are two decorrelated sources of uncertainties. 
Conversely, at fixed energy $\Delta_\sigma^\mathrm{net}$ decreases by increasing $N_\rho$ and/or $N_\mathrm{n}$, thus confirming our working hypothesis that an infinitely large neural network, trained over an infinitely large dataset, is able to provide the exact answer independently of the initialization parameters. 
$\Delta_\sigma^\mathrm{net}$ is instead stable with respect to changes in the dimension $N_\mathrm{b}$ of the functional-basis. This stability can be ascribed to the fact that we are targeting the extraction of smeared spectral densities at sufficiently large values of $\sigma$ so that the local complexity introduced in the unsmeared spectral densities by increasing $N_\mathrm{b}$ is washed out. A trend in $\Delta_\sigma^\mathrm{net}$ as a function of $N_\mathrm{b}$ may thus become visible when considering smaller values of $\sigma$.

\bibliography{main}

\end{document}